\documentclass[lettersize,journal]{IEEEtran}
\usepackage{amsmath,amsfonts}
\usepackage[noend]{algpseudocode} 
\usepackage{algorithm} 
\usepackage{amsmath}
\usepackage{array}
\usepackage[caption=false,font=normalsize,labelfont=sf,textfont=sf]{subfig}
\usepackage{textcomp}
\usepackage{stfloats}
\usepackage{url}
\usepackage{verbatim}
\usepackage{graphicx}
\usepackage{subcaption}

\def\BibTeX{{\rm B\kern-.05em{\sc i\kern-.025emb}\kern-.08emT\kern-.1667em\lower.7ex\hbox{E}\kern-.125emX}}
\usepackage{balance}
\begin{document}
\title{Two-stage Coordinated Energy Management of Train Operation and Wayside Energy Storage System for Rail Power Supply Systems}

\author{Fei Liu, \textit{Graduate Student Member, IEEE}, Niklas Biedermann, Stefan Östlund, \textit{Senior Member, IEEE}, Qianwen Xu, \textit{Senior Member, IEEE}
\thanks{This work was supported by Europe’s Rail Flagship Project 4 - Sustainable and green rail systems, and STandUP for Energy.

This work has been submitted to the IEEE for possible publication. Copyright may be transferred without notice, after which this version may no longer be accessible.}}

\markboth{Journal of \LaTeX\ Class Files,~Vol.~, No.~, Month~Year}%
{How to Use the IEEEtran \LaTeX \ Templates}

\maketitle

\begin{abstract}

The increasing electrification of railway power supply system (RPSS) intensify the operational and economic challenges at the railway power system interface. Energy storage systems (ESSs) can provide fast and flexible support to mitigate short term power spikes and to improve the energy management. 
However, achieving coordinated operation is challenged by the tight coupling among electrical railway operation and ESS dispatch under time-varying traction demand and network limits.
This paper proposes a two-stage coordinated energy management method for electrified RPSSs that jointly optimizes railway system operation, train trajectories and ESS dispatch while explicitly accounting for traction power flow constraints.
First, a day-ahead operation stage determines the train operating profiles and the ESS setting decisions to establish the baseline operating plan. Then, an intra-day rolling optimization stage based on adaptive weight economic-model predictive control (AWC-MPC) updates ESS dispatch under refreshed forecasts of traction demand and renewable output. A real Swedish railway case is utilized to minimize the energy purchase cost and the ESS cost while reducing peak grid power, demonstrating its practical applicability with 35.7\% peak grid power demand and 28.8\% total system cost reduction.

\end{abstract}


\section{Introduction}
Railway systems are an important component of modern transportation infrastructure, offering safety, high capacity, and sustainability~\cite{b1}. In pursuit of climate neutrality goals, the European Union has prioritized railway decarbonization through programs such as Horizon Europe and the EU-Rail initiative~\cite{b2,b3}. These efforts accelerate the deployment of advanced energy management technologies with the energy storage systems (ESSs) integration. The aim is to enhance energy efficiency, integrate renewable energy resources (RESs) and support the shift toward low-carbon railway transportation~\cite{a1}. With the growing global urgency to boost energy efficiency and cut carbon emissions, the railway sector is placing greater emphasis on advanced energy management and system optimization. Moreover, as the railway traction network becomes more tightly coupled with the grid, railway energy use and operating costs are increasingly influenced by grid side conditions, so the traction system can no longer be treated as an isolated, passive electricity consumer, which calls for integrated grid railway planning~\cite{a2}.

To build a more sustainable transport system, railway networks must meet the increasing energy demands while ensuring operational resilience. 
In conventional railway power supply systems (RPSSs)~\cite{a3}, traction transformers are fed directly from the public grid and deliver one directional power to the catenary, and the regenerative braking energy (RBE) is largely dissipated in onboard resistors. With the integration of ESSs and RESs, this RPSS architecture is coordinated with many power conditioning converters, which enables bidirectional power exchange, recovery of RBE, peak shaving, and voltage regulation~\cite{b7,b8}, thereby improving energy efficiency and mitigating the impact of traction loads on the grid~\cite{a4}.
Among various ESS configurations, onboard systems support individual trains by directly reusing recovered energy for acceleration and load smoothing~\cite{b9,b10}, while wayside ESS enables energy redistribution across multiple trains at the network level. The latter offers advantages in scalability and interaction with the grid~\cite{b11}, and is therefore considered a practical and cost-effective solution for enhancing energy efficiency.

Recent research has explored diverse approaches to improve the performance of ESSs in transport applications. Thermal-constrained control with deep Q-learning for co-phase traction systems balances power flow and suppresses peaks while satisfying device temperature limits~\cite{b12}. Operation-aware coordination that reuses RBE to fast charging stations improves load balancing and strengthens network stability~\cite{b13}. In rail applications, multi-application schemes that combine static power conditioners with storage maximize peak clipping and regenerative energy capture, delivering measurable economic gains~\cite{b14}. Extending these ideas across multiple time scales, a hierarchical framework for urban rail with distributed photovoltaic(PV) and hybrid storage is designed, which coordinates multiple control stages to raise PV and RBE utilization and to cut operating and carbon-trading costs~\cite{a5}. 

However, most existing works either focus solely on peak shaving or energy recovery, without jointly optimizing train operation within an integrated railway–power system model. Some recent studies have attempted to bridge this gap by considering both trajectory planning and energy aspects. A two-step method is developed in~\cite{b15} that jointly optimizes train operation, timetable, and the management of energy storage devices, but it only considers onboard ESS without coordinating with wayside storage. In~\cite{b16}, authors introduced a Train–Network–HESS integrated model that co-optimizes train trajectories and wayside hybrid ESS configuration. Nevertheless, its power flow treatment is confined to single-point substations. In parallel, cost-oriented sizing studies place PVs, winds, and ESSs along AC railways while assuming fixed train demand and without modeling grid rail coordination beyond the traction network~\cite{a6}, and network level analyses quantify PVs and ESSs impacts on voltages, currents and conductor temperatures but do not couple investment and operation in a unified optimization~\cite{a7}. Furthermore, many studies rely on  simplified railway representations, which omit real topology, equipment ratings, and operational or protection limits, making the results difficult to apply in practice and offering limited guidance for planning and operation in real railway systems.

To address these challenges, this paper proposes a two-stage coordinated optimization method for electrified RPSS. The method dynamically couples train movement control with the siting and sizing of wayside ESSs installations, which aims to reduce peak grid power demand and minimize the overall system cost of electrified railway traction. The main contributions of this study are summarized as follows: 

1) A high-fidelity RPSS model is developed by integrating multi-train movement and trajectory reference generation with electrical railway, power flow constraints, timetable constraints, and electricity cost with time-varying prices and peak demand limits.

2) A two-stage coordinated optimization model for RPSS energy management is proposed. In Stage I, a day-ahead operation stage determines wayside ESS siting and sizing together with train trajectory, and produce the grid power and ESS operation references. In Stage II, intra-day rolling optimization based on adaptive weight economic-model predictive control (AWC-MPC) is implemented to update ESSs charging and discharging commands using refreshed forecasts of traction demand and PV power.

3) A Swedish railway system map and train timetable are utilized to capture the interactions between train operations, ESSs and RESs behavior, power flow analysis, and grid conditions. Peak grid power demand reduction and total system cost reduction under realistic operating conditions are demonstrated.

The remainder of this paper is structured as follows.
Section II describes the RPSS-grid integration model, including high fidelity train movement model, railway power supply and power flow calculations.
Section III presents the two-stage coordinated optimization model that integrates ESSs placement and train operation.
Section IV provides case studies based on a Swedish railway network to validate the proposed approach.
Finally, Section V concludes the paper.

\section{Structure of RPSS model}

\begin{figure}[htbp]
\centerline{\includegraphics[width=0.4\textwidth]{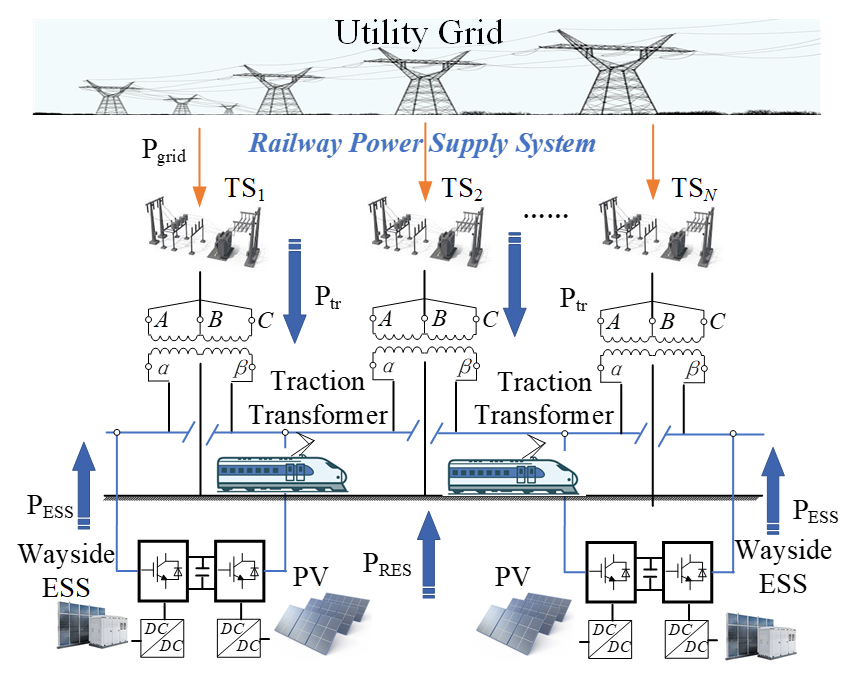}}
\caption{Structure diagram of the RPSS model including train and ESS integration.}
\label{fig:FTPSS_Grid_structure}
\end{figure}

In this section, a physics-based train movement model is developed to simulate the operational characteristics of multi-trains on an electrified railway network. In addition, the RPSS model characterizes the electrical interactions among traction substations, section posts, and line segments, and interfaces RESs and ESSs as nodal power injections. Fig.~\ref{fig:FTPSS_Grid_structure} shows the structure of the RPSS model.

\subsection{Multi-Train Movement Modeling}

To support coordinated operation and subsequent optimization, this subsection presents a multi-train interaction description and a trajectory reference generation based on the train movement model. The model outputs time-stamped spatiotemporal states of each train, including variables $(s_h(t), v_h(t), P_h(t), Q_h(t))$, which are used to construct time-varying traction power demands for traction power flow and grid interaction. $s_h(t)$ is the position of train $h$ along the track at time $t$, $v_h(t)$ is the speed, and $P_h(t)$ and $Q_h(t)$ are the instantaneous active and reactive power demands.

\subsubsection{Single-Train Dynamics Model}

The movement of a railway train is constrained by journey gradient profiles and speed limits. The dynamic behavior of train $h \in \mathcal{H}$ can be described as:
\begin{equation}
M'_h \frac{dv_h}{dt} = F^{h}_{\mathrm{tr}}(t) - F^{h}_{\mathrm{res}}(v_h) - F^{h}_{\mathrm{sl}}(s_h),
\label{eq:dyn_time}
\end{equation}
where $M'_h$ is the effective mass of train $h$ including rotary allowance, and $F^{h}_{\mathrm{tr}}(t)$ is the traction or braking force.

The resistance and slope components are expressed as:
\begin{align}
F^{h}_{\mathrm{res}}(v_h) &= A_h + B_h v_h + C_h v_h^2, \label{eq:davis_ext}\\
F^{h}_{\mathrm{sl}}(s_h) &= M' g \sin(\alpha(s_h)), \label{eq:slope_ext}
\end{align}
where $M'$ is the tare mass of train, $F^{h}_{\mathrm{res}}(v_h)$ is the resistance described by Davis coefficients $A_h$, $B_h$, and $C_h$, $F^{h}_{\mathrm{sl}}(s_h)$ is the slope resistance, $g$ is the gravitational acceleration, and $\alpha(\cdot)$ is the track slope angle.

To better handle position-dependent constraints such as gradient and speed limits, eq. \eqref{eq:dyn_time} can be equivalently reformulated in the space domain for $v_h(s)>0$:
\begin{align}
\frac{dv_h(s)}{ds} &= \frac{F^{h}_{\mathrm{tr}}(s) - F^{h}_{\mathrm{res}}(v_h(s)) - F^{h}_{\mathrm{sl}}(s)}{M'_h\, v_h(s)},
\label{eq:dyn_space}\\
\frac{dt_h(s)}{ds} &= \frac{1}{v_h(s)},
\label{eq:t_space}
\end{align}
where $v_h(s)$ is the speed as a function of position $s$ and $t_h(s)$ denotes the travel time mapping along the route. 

\subsubsection{Trajectory Reference Construction}

Let the route be discretized into $N$ spatial nodes $\{s_q\}_{q=0}^{N}$, where $s_0$ and $s_N$ denote the departure and arrival positions. A reference trajectory is defined by a speed vector:
\begin{equation}
\mathbf{v}^{\mathrm{ref}}_h = \left[v^{\mathrm{ref}}_{h,0}, v^{\mathrm{ref}}_{h,1}, \ldots, v^{\mathrm{ref}}_{h,N}\right]^{\top},
\end{equation}
which should satisfy the boundary and operational constraints:
\begin{align}
0 \le v^{\mathrm{ref}}_{h,q} &\le v_{\lim}(s_q), \quad \forall q,\\
a_{\min} \le \frac{\left(v^{\mathrm{ref}}_{h,q}\right)^2-\left(v^{\mathrm{ref}}_{h,q-1}\right)^2}{2\Delta s_q} &\le a_{\max}, \quad \forall q\ge 1,
\label{eq:acc_constraint_ref}
\end{align}
where $v_{\lim}(s_q)$ is the speed limit at position $s_q$, $a_{\min}$ and $a_{\max}$ denote the minimum and maximum allowable accelerations, and $\Delta s_q = s_q - s_{q-1}$.

Given $(s_{q-1}, v^{\mathrm{ref}}_{h,q-1})$ and $(s_q, v^{\mathrm{ref}}_{h,q})$, the corresponding time stamps can be obtained by the time-space mapping, which enables reconstructing the continuous time reference $(s_h^{\mathrm{ref}}(t), v_h^{\mathrm{ref}}(t))$ via piecewise linear interpolation:
\begin{equation}
t^{\mathrm{ref}}_{h,q} = t^{\mathrm{ref}}_{h,q-1} + \frac{2\Delta s_q}{v^{\mathrm{ref}}_{h,q-1}+v^{\mathrm{ref}}_{h,q}},
\label{eq:time_space_map}
\end{equation}

\subsubsection{Power Demand Calculation}

At each time instant $t$, the mechanical power associated with the traction force is $P^{h}_{\mathrm{mech}}(t)=F^{h}_{\mathrm{tr}}(t)\, v_h(t)$. The active power demand is modeled as:
\begin{equation}
P_h(t)=
\begin{cases}
\displaystyle \frac{P^{h}_{\mathrm{mech}}(t)}{\eta^{h}_{\mathrm{tr}}(v_h,F^{h}_{\mathrm{tr}})} + P^{h}_{\mathrm{aux}}, & P^{h}_{\mathrm{mech}}(t)\ge 0,\\[6pt]
\displaystyle -\eta^{h}_{\mathrm{reg}}(v_h,F^{h}_{\mathrm{tr}})\, \bigl|P^{h}_{\mathrm{mech}}(t)\bigr| + P^{h}_{\mathrm{aux}}, & P^{h}_{\mathrm{mech}}(t)<0,
\end{cases}
\label{eq:power_eff}
\end{equation}
where $\eta^{h}_{\mathrm{tr}}(\cdot)$ and $\eta^{h}_{\mathrm{reg}}(\cdot)$ denote traction and regenerative conversion efficiencies, respectively, and $P^{h}_{\mathrm{aux}}$ is the auxiliary power. The reactive power can be described by:
\begin{equation}
Q_h(t)= P_h(t)\tan\varphi_h + Q^{h}_{\mathrm{aux}},
\label{eq:reactive_ext}
\end{equation}
where $\varphi_h$ is the phase angle related to the train power factor and $Q^{h}_{\mathrm{aux}}$ is the auxiliary reactive demand.

\subsubsection{Multi-Train Spatiotemporal Interaction}

The spatiotemporal state of each train is generated by simulating eq. \eqref{eq:dyn_space}. For dense traffic, neighboring trains in the same direction should satisfy a safe spacing constraint for all time step $t$:
\begin{equation}
s_h(t)-s_{h-1}(t)\ge d_{\mathrm{safe}},
\label{eq:safe_spacing}
\end{equation}
where $d_{\mathrm{safe}}$ is a speed dependent margin to represent signaling and braking safety requirements.

\subsection{RPSS Modeling with RESs and ESSs integration}

This subsection models the railway power supply and demand on the grid side using an equivalent injection representation, which provides a direct interface between railway operation states and the network constraints used in the subsequent optimization.

\subsubsection{Electrified Railway System Modelling}

The electrified railway system is represented by a graph:
\begin{equation}
\mathcal{G}_{\mathrm{G}}=(\mathcal{N}_{\mathrm{G}},\mathcal{E}_{\mathrm{G}}),
\end{equation}
where $\mathcal{N}_{\mathrm{G}}$ is the set of buses and $\mathcal{E}_{\mathrm{G}}$ is the set of branches.

Let $\mathcal{N}\subseteq\mathcal{N}_{\mathrm{G}}$ denote the set of buses associated with railway supply and load representation:
\begin{equation}
\mathcal{N}=\mathcal{N}_{\mathrm{TS}}\cup\mathcal{N}_{\mathrm{SP}},
\end{equation}
where $\mathcal{N}_{\mathrm{TS}}$ is the set of traction substation buses, $\mathcal{N}_{\mathrm{SP}}$ is the set of section-post buses used for geographical segmentation and load mapping.

At each bus $b\in\mathcal{N}$ and time step $t$, the railway-related active and reactive injections are decomposed as:
\begin{align}
P_b^{\mathrm{R}}(t) &= P_b^{\mathrm{RES}}(t)+P_b^{\mathrm{ESS}}(t)-P_b^{\mathrm{tr}}(t),
\label{eq:rail_P_decomp}\\
Q_b^{\mathrm{R}}(t) &= Q_b^{\mathrm{RES}}(t)+Q_b^{\mathrm{ESS}}(t)-Q_b^{\mathrm{tr}}(t),
\label{eq:rail_Q_decomp}
\end{align}
where $P_b^{\mathrm{R}}(t)$ and $Q_b^{\mathrm{R}}(t)$ are the net railway-related active and reactive injections at bus $b$,
$P_b^{\mathrm{RES}}(t)$ and $Q_b^{\mathrm{RES}}(t)$ denote the PV injections,
$P_b^{\mathrm{ESS}}(t)$ and $Q_b^{\mathrm{ESS}}(t)$ denote the ESS injections,
and $P_b^{\mathrm{tr}}(t)$ and $Q_b^{\mathrm{tr}}(t)$ denote the mapped train active and reactive powers at bus $b$.

\subsubsection{Multi-Train Load Mapping to Grid Buses}

Using the train power profiles $P_h(t)$ and $Q_h(t)$ obtained from the train movement model, the mapped traction demands are constructed as:
\begin{align}
P_b^{\mathrm{tr}}(t) &= \sum_{h\in\mathcal{H}} \pi_{h,b}(t)\, P_h(t),
\quad \forall b\in\mathcal{N},\ \forall t,
\label{eq:train_map_gridP}\\
Q_b^{\mathrm{tr}}(t) &= \sum_{h\in\mathcal{H}} \pi_{h,b}(t)\, Q_h(t),
\quad \forall b\in\mathcal{N},\ \forall t.
\label{eq:train_map_gridQ}
\end{align}
where $\mathcal{H}$ is the set of trains, $\pi_{h,b}(t)$ is the allocation factor from train $h$ to bus $b$ satisfying $0\le \pi_{h,b}(t)\le 1$ and $\sum_{b\in\mathcal{N}}\pi_{h,b}(t)=1$, and $t$ is the discrete time step index.

\section{Two-Stage Coordinated Optimization Model}

This section proposes a coordinated optimization model that couples train operation decisions with wayside ESS planning under RPSS operating constraints. The following subsections formulate solution procedure first, and then provides two-stage optimization explicitly.

\subsection{Solution of the two-stage coordinated optimization model for RPSS energy management}

\begin{figure}[htbp]
\centerline{\includegraphics[width=0.45\textwidth]{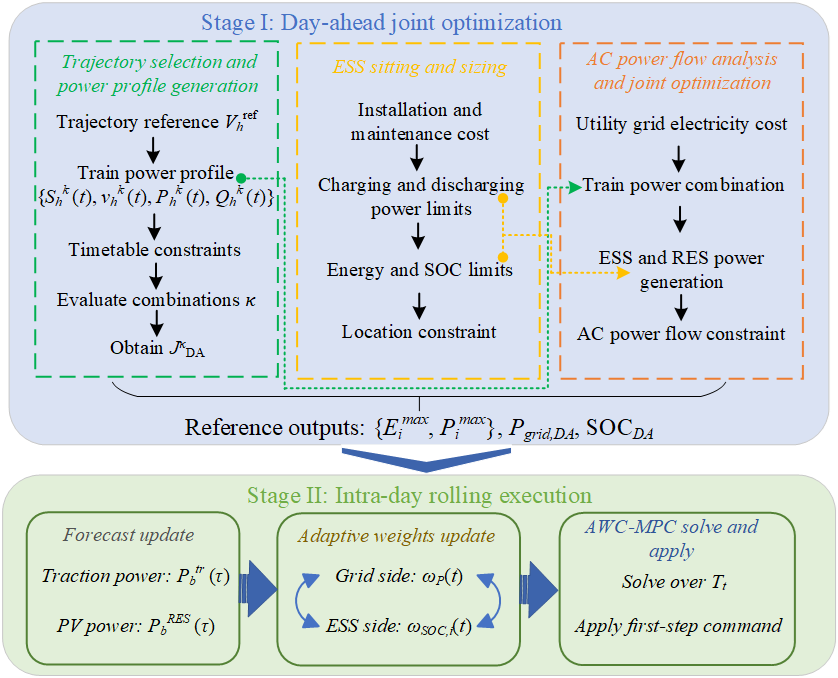}}
\caption{Two-stage coordinated optimization model
for RPSS energy management.}
\label{fig:wholemethod}
\end{figure}

This subsection proposes a six-step solution procedure for two-stage coordination optimization, shown in Fig.~\ref{fig:wholemethod}. Stage~I determines the baseline train trajectories and the ESS siting and sizing decisions while enforcing the coupled grid constraints. Based on the day-ahead plan, Stage~II performs a receding horizon optimization to mitigate deviations caused by short term variations in traction demand and PV output, by updating the ESS charging and discharging commands to track the day-ahead references of $P_{\mathrm{grid}}(\tau)$ and $\mathrm{SOC}_i(\tau)$.

\textit{Step 1: Day-ahead data preparation and load mapping}: Using the candidate train trajectories, the corresponding traction power profiles are generated by the train movement model and mapped to grid buses. The resulting railway power injections are constructed, together with the day-ahead PV forecasts and electricity price $\rho(t)$ over $t\in\mathcal{T}_{\mathrm{DA}}$.

\textit{Step 2: Trajectory selection coupled with ESS planning}: For each candidate trajectory combination indexed by $\kappa$, the day-ahead operation model \eqref{eq:DA_obj_total} is solved under the resulting railway injections together with the grid and ESS constraints, yielding an optimal objective value denoted by $J_{\mathrm{DA}}^{\kappa}$. The baseline trajectory combination is selected by minimizing $J_{\mathrm{DA}}^{\kappa}$ over the candidate combinations.

\textit{Step 3: Stage~I day-ahead joint optimization}: Solving the day-ahead operation problem subject to the trajectory feasibility constraints, the ESS operation constraints, and the grid constraints, Stage~I outputs the ESS installation decisions and ratings $\{E_i^{\max},P_i^{\max}\}$, as well as the day-ahead reference trajectories $\{P_{\mathrm{grid,DA}}(t),\mathrm{SOC}_{i,\mathrm{DA}}(t)\}$.

\textit{Step 4: Intra-day forecast update and state obtain}: At each real-time step $t$, Stage~II constructs the prediction horizon $\mathcal{T}_t=\{t,t+1,\ldots,t+H-1\}$ using the latest forecasts of $\{P_b^{\mathrm{tr}}(\tau),P_b^{\mathrm{RES}}(\tau)\}_{\tau\in\mathcal{T}_t}$ and acquires the current ESS states $\{\mathrm{SOC}_i(t)\}_{i\in\mathcal{N}_{\mathrm{sta}}}$.

\textit{Step 5: Adaptive weight update and AWC-MPC optimization}: Given the operating conditions at time step $t$, the tracking weights are updated using the electricity price $\rho(t)$, the current SOC states $\mathrm{SOC}_i(t)$, and the PV uncertainty indicator $\sigma_{\mathrm{pv}}(t)$. Stage~II solves the AWC-MPC problem \eqref{eq:awc_mpc_obj} over $\mathcal{T}_t$ to obtain the updated ESS charging and discharging commands $\{P_i^{\mathrm{ESS,c}}(\tau),P_i^{\mathrm{ESS,d}}(\tau)\}_{\tau\in\mathcal{T}_t}$.

\textit{Step 6: Receding horizon implementation over the operating horizon}: The resulting grid exchange $P_{\mathrm{grid}}(\tau)$ is evaluated according to \eqref{eq:RT_Pgrid_balance}. Only the first-step commands at $\tau=t$ are applied to the real-time control layer, and the horizon is shifted forward by one step for the next iteration at $t+1$. The intra-day horizon continuously corrects day-ahead deviations while tracking the planned $P_{\mathrm{grid,DA}}(\tau)$ and $\mathrm{SOC}_{i,\mathrm{DA}}(\tau)$ trajectories under updated forecasts.

\subsection{Stage I: Day-Ahead Operation With Trajectory Selection and ESS Siting and Sizing}

\subsubsection{Objective Function}

The Stage I objective minimizes the total day-ahead cost:
\begin{equation}
\min\ J_{\mathrm{DA}} = C_{\mathrm{grid}} + C_{\mathrm{ESS}},
\label{eq:DA_obj_total}
\end{equation}
where $C_{\mathrm{grid}}$ denotes the grid purchase cost including a peak penalty and $C_{\mathrm{ESS}}$ denotes the annualized ESS investment cost allocated to the day-ahead horizon.

The grid purchase cost is defined as:
\begin{align}
C_{\mathrm{grid}}
&=\sum_{t\in\mathcal{T}_{\mathrm{DA}}}\rho(t)\,P_{\mathrm{grid}}(t)\,\Delta t
+\lambda_{\mathrm{G,peak}} P_{\mathrm{G}}^{\mathrm{peak}},
\label{eq:C_grid}
\end{align}
where $\rho(t)$ is the electricity price at time step $t$, $P_{\mathrm{grid}}(t)$ is the aggregated active power, $\lambda_{\mathrm{G,peak}}$ is the peak penalty coefficient, and $P_{\mathrm{G}}^{\mathrm{peak}}$ is an auxiliary variable that upper-bounds the grid exchange peak.

The ESS cost is modeled as:

\begin{equation}
\begin{aligned}
C_{\mathrm{ESS}}
&=\left(\sum_{i\in\mathcal{N}_{\mathrm{sta}}}\left(c_E E_i^{\max}+c_P P_i^{\max}\right)\right)
\cdot \mathrm{CRF}\\
&\quad \cdot \frac{T}{T_{\mathrm{day}}}\cdot \left(1+c_{\mathrm{opex}}\right),
\end{aligned}
\label{eq:C_ESS}
\end{equation}
where $\mathcal{N}_{\mathrm{sta}}\subseteq\mathcal{N}$ is the set of candidate buses for ESS installation, $E_i^{\max}$ and $P_i^{\max}$ are the energy capacity and power rating at bus $i$, $c_E$ and $c_P$ are the corresponding unit investment costs, $\mathrm{CRF}$ is the capital recovery factor, $c_{\mathrm{opex}}$ is the operation and maintenance cost ratio, $T=|\mathcal{T}_{\mathrm{DA}}|$ is the number of day-ahead time steps, and $T_{\mathrm{day}}$ is the number of time steps per day under the same time resolution.

\subsubsection{Constraints}

\paragraph{Trajectory selection and train power profiles}

In Stage~I, each train $h\in\mathcal{H}$ is associated with a pre-generated set of feasible reference trajectories introduced in the train movement modeling subsection. For each candidate trajectory combination indexed by $\kappa$, the train movement model produces the corresponding time-stamped profiles $\{s_h^{\kappa}(t),v_h^{\kappa}(t),P_h^{\kappa}(t),Q_h^{\kappa}(t)\}$, which are mapped to grid buses through \eqref{eq:train_map_gridP}-\eqref{eq:train_map_gridQ} form the railway injections in \eqref{eq:rail_P_decomp}-\eqref{eq:rail_Q_decomp}. 

The day-ahead operation model \eqref{eq:DA_obj_total} is solved under the resulting railway injections together with the grid and ESS constraints, with an optimal objective value denoted by $J_{\mathrm{DA}}^{\kappa}$. The baseline trajectory combination is selected by:
\begin{equation}
\kappa^{\star}=\arg\min_{\kappa}\ J_{\mathrm{DA}}^{\kappa},
\label{eq:traj_select_argmin}
\end{equation}

\paragraph{Timetable consistency and dwell constraints}
For each train $h\in\mathcal{H}$ and station $k\in\mathcal{K}_h$, the arrival and departure instants
satisfy:
\begin{equation}
t_{h,k}^{\mathrm{dep}} = t_{h,k}^{\mathrm{arr}} + T_{h,k}^{\mathrm{dw}},
\quad \forall h\in\mathcal{H},\ \forall k\in\mathcal{K}_h,
\label{eq:DA_dwell_def}
\end{equation}
\begin{equation}
\underline{T}_{k}^{\mathrm{dw}} \le T_{h,k}^{\mathrm{dw}} \le \overline{T}_{k}^{\mathrm{dw}},
\quad \forall h\in\mathcal{H},\ \forall k\in\mathcal{K}_h,
\label{eq:DA_dwell_bounds}
\end{equation}
where $t_{h,k}^{\mathrm{arr}}$ and $t_{h,k}^{\mathrm{dep}}$ denote the arrival and departure instants of train $h$
at station $k$, $T_{h,k}^{\mathrm{dw}}$ is the dwell time, and $\underline{T}_{k}^{\mathrm{dw}}$ and
$\overline{T}_{k}^{\mathrm{dw}}$ are the minimum and maximum dwell time limits at station $k$, respectively.

\paragraph{ESS siting, sizing, and interface capability}
An ESS installation decision is represented by a binary variable $x_i\in\{0,1\}$ for each $i\in\mathcal{N}_{\mathrm{sta}}$. The sizing variables satisfy:
\begin{align}
0 \le E_i^{\max} &\le x_i \bar{E}_i,
\quad \forall i\in\mathcal{N}_{\mathrm{sta}},
\label{eq:DA_size_E}\\
0 \le P_i^{\max} &\le x_i \bar{P}_i,
\quad \forall i\in\mathcal{N}_{\mathrm{sta}},
\label{eq:DA_size_P}
\end{align}
where $\bar{E}_i$ and $\bar{P}_i$ are the upper bounds of candidate installations.

Charging and discharging powers are introduced as:
\begin{equation}
P_i^{\mathrm{ESS}}(t)=P_i^{\mathrm{ESS,d}}(t)-P_i^{\mathrm{ESS,c}}(t),
\quad \forall i\in\mathcal{N}_{\mathrm{sta}},\ \forall t\in\mathcal{T}_{\mathrm{DA}},
\label{eq:DA_ess_netP}
\end{equation}
where $P_i^{\mathrm{ESS,c}}(t)\ge 0$ and $P_i^{\mathrm{ESS,d}}(t)\ge 0$ denote charging and discharging powers.

The SOC dynamics constraints are:
\begin{align}
\mathrm{SOC}_i(t+1)
&=\mathrm{SOC}_i(t)
+\kappa_i^{\mathrm{c}} P_i^{\mathrm{ESS,c}}(t)
-\kappa_i^{\mathrm{d}} P_i^{\mathrm{ESS,d}}(t),
\nonumber\\
&\qquad \forall i\in\mathcal{N}_{\mathrm{sta}},\ \forall t\in\mathcal{T}_{\mathrm{DA}},
\label{eq:DA_soc}
\end{align}
where $\mathrm{SOC}_i(t)$ is the state of charge, and $\eta_i^{\mathrm{c}}$ and $\eta_i^{\mathrm{d}}$ are the charging and discharging efficiencies. Efficiency parameters are $\kappa_i^{\mathrm{c}}=\eta_i^{\mathrm{c}}\Delta t/E_i^{\max}$ and $\kappa_i^{\mathrm{d}}=\Delta t/(\eta_i^{\mathrm{d}}E_i^{\max})$. 

The converter capability constraints are:
\begin{equation}
0 \le P_i^{\mathrm{ESS,c}}(t), P_i^{\mathrm{ESS,d}}(t) \le P_i^{\max},
\label{eq:DA_ess_ch}
\end{equation}

\begin{equation}
\begin{aligned}
\left(P_i^{\mathrm{ESS}}(t)\right)^2+\left(Q_i^{\mathrm{ESS}}(t)\right)^2
&\le \left(S_i^{\max}\right)^2,\\
&\hfill \forall i\in\mathcal{N}_{\mathrm{sta}},\ \forall t\in\mathcal{T}_{\mathrm{DA}},
\end{aligned}
\label{eq:DA_ess_S}
\end{equation}
where $Q_i^{\mathrm{ESS}}(t)$ is the reactive power injection of the ESS interface and $S_i^{\max}$ is the apparent power rating.

\paragraph{Ac power flow constraints}
For each grid bus $b\in\mathcal{N}_{\mathrm{G}}$ at time step $t$, the complex current injection is written as:
\begin{equation}
I_b(t)=\sum_{m:(b,m)\in\mathcal{E}_{\mathrm{G}}}\frac{V_b(t)-V_m(t)}{Z_{bm}},
\quad \forall b\in\mathcal{N}_{\mathrm{G}},\ \forall t\in\mathcal{T}_{\mathrm{DA}},
\label{eq:Iz_compact}
\end{equation}
where $V_b(t)$ is the complex voltage at bus $b$, $I_b(t)$ is the complex current injection,
$Z_{bm}=R_{bm}+\mathrm{j}X_{bm}$ is the line impedance of branch $(b,m)$,
$R_{bm}$ and $X_{bm}$ are the resistance and reactance, respectively.

The nodal active and reactive power injections are obtained from:
\begin{equation}
P_b^{\mathrm{R}}(t)+\mathrm{j}Q_b^{\mathrm{R}}(t)=V_b(t)\,I_b^{*}(t),
\quad \forall b\in\mathcal{N}_{\mathrm{G}},\ \forall t\in\mathcal{T}_{\mathrm{DA}},
\label{eq:PQ_from_VI}
\end{equation}
where $P_b^{\mathrm{R}}(t)$ and $Q_b^{\mathrm{R}}(t)$ are the railway related injections defined in
\eqref{eq:rail_P_decomp}--\eqref{eq:rail_Q_decomp}, and $(\cdot)^*$ denotes the complex conjugate.

\subsection{Stage II: Intra-Day Rolling Optimization Based on AWC-MPC}

\subsubsection{AWC-MPC Objective Function}

Define the output vector:
\begin{equation}
\mathbf{y}(t)=
\begin{bmatrix}
P_{\mathrm{grid}}(t)\\
\mathbf{SOC}(t)
\end{bmatrix},
\label{eq:y_vec}
\end{equation}
where $\mathbf{SOC}(t)=\left[\mathrm{SOC}_i(t)\right]_{i\in\mathcal{N}_{\mathrm{sta}}}$ collects the SOC states of installed ESS units, and let $\mathbf{y}_{\mathrm{DA}}(t)$ denote the corresponding day-ahead reference vector obtained in Stage I. Over a prediction horizon $\mathcal{T}_t=\{t,t+1,\ldots,t+H-1\}$, the AWC-MPC objective is formulated as:
\begin{align}
\min\ J_{\mathrm{RT}}(t)
&=\sum_{\tau\in\mathcal{T}_t}
\left(\mathbf{y}(\tau)-\mathbf{y}_{\mathrm{DA}}(\tau)\right)^{\top}
\mathbf{W}(t)
\left(\mathbf{y}(\tau)-\mathbf{y}_{\mathrm{DA}}(\tau)\right)
\nonumber\\
&\quad+\left(\Delta\mathbf{u}(t)\right)^{\top}\mathbf{Q}\,\Delta\mathbf{u}(t),
\label{eq:awc_mpc_obj}
\end{align}
where $\mathbf{W}(t)$ is an adaptive weighting matrix updated at time step $t$, $\mathbf{Q}$ is a diagonal smoothing matrix consistent with the weights, $\mathbf{u}(t)$ is the control vector over the prediction horizon, and $\Delta\mathbf{u}(t)$ is the corresponding control increment vector.

To match the tracking and smoothing interpretation, \eqref{eq:awc_mpc_obj} can be written in an equivalent scalar form as:
\begin{align}
\ J_{\mathrm{RT}}(t) &=
\nonumber\\
&\sum_{\tau\in\mathcal{T}_t}
\omega_{P}(t)\left(P_{\mathrm{grid}}(\tau)-P_{\mathrm{grid,DA}}(\tau)\right)^2
\nonumber\\
&\quad+\sum_{\tau\in\mathcal{T}_t}\sum_{i\in\mathcal{N}_{\mathrm{sta}}}
\omega_{\mathrm{SOC},i}(t)
\left(\mathrm{SOC}_i(\tau)-\mathrm{SOC}_{i,\mathrm{DA}}(\tau)\right)^2
\nonumber\\
&\quad+\sum_{\tau\in\mathcal{T}_t\setminus\{t\}}\sum_{i\in\mathcal{N}_{\mathrm{sta}}}
\Bigl(
q_{i}^{\mathrm{d}}\left(\Delta P_{i}^{\mathrm{ESS,d}}(\tau)\right)^2
\nonumber\\
&\qquad\qquad\qquad
+q_{i}^{\mathrm{c}}\left(\Delta P_{i}^{\mathrm{ESS,c}}(\tau)\right)^2
\Bigr),
\label{eq:awc_mpc_scalar}
\end{align}
where $P_{\mathrm{grid,DA}}(\tau)$ and $\mathrm{SOC}_{i,\mathrm{DA}}(\tau)$ are the day-ahead references of the
grid exchange and the SOC of ESS $i$, respectively. $\omega_{P}(t)$ and $\omega_{\mathrm{SOC},i}(t)$ are the adaptive tracking weights. $q_{i}^{\mathrm{d}}$ and $q_{i}^{\mathrm{c}}$ are the smoothing weights for the
discharging and charging actions, respectively, and $\Delta P_{i}^{\mathrm{ESS,d}}(\tau)$ and
$\Delta P_{i}^{\mathrm{ESS,c}}(\tau)$ denote the control increments defined in
\eqref{eq:du_dis_def}--\eqref{eq:du_ch_def}.

The adaptive weights are updated according to operating conditions:
\begin{align}
\omega_{P}(t) &= \omega_{P}^{0}\, \phi_{\rho}\bigl(\rho(t)\bigr),
\label{eq:wP_update}\\
\omega_{\mathrm{SOC},i}(t) &= \omega_{\mathrm{SOC}}^{0}\, \phi_{\mathrm{soc}}\bigl(\mathrm{SOC}_i(t)\bigr)\,\phi_{\sigma}\bigl(\sigma_{\mathrm{pv}}(t)\bigr),
\label{eq:wSOC_update}
\end{align}
where $\rho(t)$ is the electricity price, $\sigma_{\mathrm{pv}}(t)$ is a PV uncertainty indicator, $\omega_{P}^{0}$ and $\omega_{\mathrm{SOC}}^{0}$ are base weights, and $\phi_{\rho}(\cdot)$, $\phi_{\mathrm{soc}}(\cdot)$, and $\phi_{\sigma}(\cdot)$ are scaling functions. When the electricity price $\rho(t)$ is high, $\phi_{\rho}\!\left(\rho(t)\right)$ increases $\omega_{P}(t)$ to prioritize tracking of $P_{\mathrm{grid}}(\tau)$ to the day-ahead reference. Meanwhile, $\omega_{\mathrm{SOC},i}(t)$ is amplified through $\phi_{\mathrm{soc}}\!\left(\mathrm{SOC}_i(t)\right)$ and $\phi_{\sigma}\!\left(\sigma_{\mathrm{pv}}(t)\right)$ when the SOC approaches its operational limits or the PV uncertainty indicator rises, so that the MPC preserves SOC margins for subsequent disturbance accommodation. In this way, the AWC mechanism enables an adaptive trade-off between economic tracking performance and operational robustness.

\subsubsection{Constraints}

\paragraph{Control vector and smoothing definition}
For each installed ESS $i\in\mathcal{N}_{\mathrm{sta}}$, define the charging and discharging control sequences
over the prediction horizon as:
\begin{equation}
\begin{aligned}
\mathbf{u}_i(t)=
\begin{bmatrix}
P_{i}^{\mathrm{ESS,c}}(t) & \cdots & P_{i}^{\mathrm{ESS,c}}(t+H-1)
\\
P_{i}^{\mathrm{ESS,d}}(t) & \cdots & P_{i}^{\mathrm{ESS,d}}(t+H-1)
\end{bmatrix}^{\top},
\end{aligned}
\label{eq:u_vec}
\end{equation}
and define the stacked control vector $\mathbf{u}(t)=\left[\mathbf{u}_i(t)\right]_{i\in\mathcal{N}_{\mathrm{sta}}}$.
The corresponding increment variables are defined element-wise as:
\begin{equation}
\begin{aligned}
\Delta P_{i}^{\mathrm{ESS,d}}(\tau)
&= P_{i}^{\mathrm{ESS,d}}(\tau)-P_{i}^{\mathrm{ESS,d}}(\tau-1),\\
&\hfill \forall i\in\mathcal{N}_{\mathrm{sta}},\ \forall \tau\in\mathcal{T}_t\setminus\{t\},
\end{aligned}
\label{eq:du_dis_def}
\end{equation}

\begin{equation}
\begin{aligned}
\Delta P_{i}^{\mathrm{ESS,c}}(\tau)
&= P_{i}^{\mathrm{ESS,c}}(\tau)-P_{i}^{\mathrm{ESS,c}}(\tau-1),\\
&\hfill \forall i\in\mathcal{N}_{\mathrm{sta}},\ \forall \tau\in\mathcal{T}_t\setminus\{t\}.
\end{aligned}
\label{eq:du_ch_def}
\end{equation}
and $\Delta\mathbf{u}(t)$ is constructed by stacking
$\{\Delta P_{i}^{\mathrm{ESS,d}}(\tau),\Delta P_{i}^{\mathrm{ESS,c}}(\tau)\}$ over $i\in\mathcal{N}_{\mathrm{sta}}$
and $\tau\in\mathcal{T}_t\setminus\{t\}$.

\paragraph{Grid power balance and receding horizon operation}
For each $\tau\in\mathcal{T}_t$, the grid exchange is defined by the active-power balance:
\begin{align}
P_{\mathrm{grid}}(\tau)
&=P_{\mathrm{grid}}^{\mathrm{base}}(\tau)
+\sum_{b\in\mathcal{N}} P_b^{\mathrm{tr}}(\tau)
-\sum_{b\in\mathcal{N}} P_b^{\mathrm{RES}}(\tau)
\nonumber\\
&\qquad-\sum_{i\in\mathcal{N}_{\mathrm{sta}}} P_i^{\mathrm{ESS}}(\tau).
\label{eq:RT_Pgrid_balance}
\end{align}
where $P_{\mathrm{grid}}^{\mathrm{base}}(\tau)$ denotes the net grid supply after serving the native loads, $P_b^{\mathrm{tr}}(\tau)$ is the mapped train active power at bus $b$, $P_b^{\mathrm{RES}}(\tau)$ is the PV injection at bus $b$, and $P_i^{\mathrm{ESS}}(\tau)$ is the ESS injection at candidate bus $i$.

At each time step $t$, the AWC-MPC problem \eqref{eq:awc_mpc_obj} is solved over $\mathcal{T}_t$ using the latest forecasts of $\{P_b^{\mathrm{tr}}(\tau),P_b^{\mathrm{RES}}(\tau)\}_{\tau\in\mathcal{T}_t}$ and the current SOC states $\{\mathrm{SOC}_i(t)\}_{i\in\mathcal{N}_{\mathrm{sta}}}$. Only the first-step updated references together with the corresponding first-step ESS commands are applied to the real-time control layer.

\section{Case Study}

\begin{figure}[htbp]
\centerline{\includegraphics[width=0.45\textwidth]{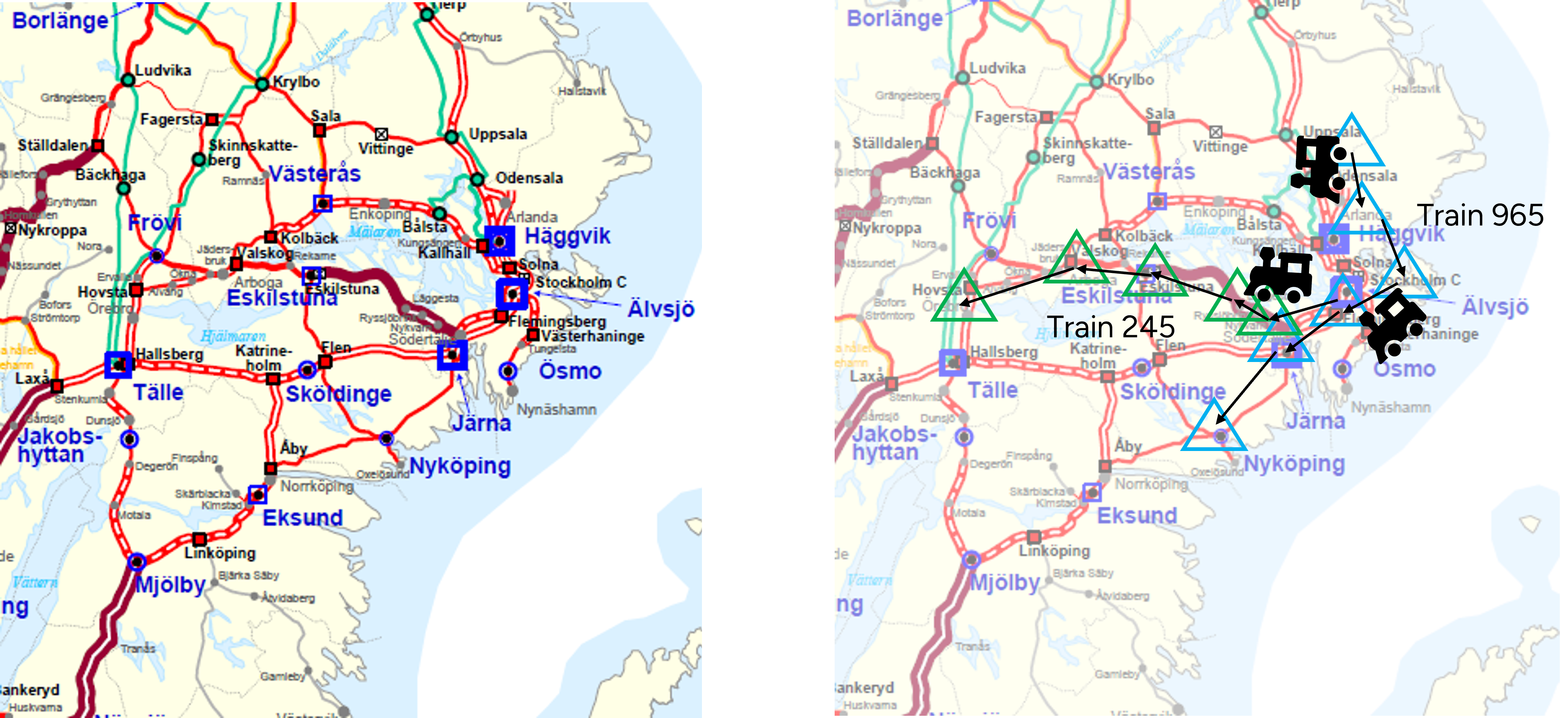}}
\caption{The Swedish railway map and train routes.}
\label{fig:sweden_map}
\end{figure}

The proposed two-stage strategy is validated on a Swedish electrified railway case study. Sweden adopts a RPSS with a nominal frequency of $16.7$~Hz and a nominal voltage level of $15$~kV (AC), which motivates coordinating train operation and track side flexibility resources under time-varying electricity prices. The study area is selected from the main corridors in the Stockholm and M\"alardalen region. The geographical locations of key railway nodes and traction supply facilities are illustrated in Fig.~\ref{fig:sweden_map}.

A pair of intersecting corridors is considered to represent realistic multi-train interaction. Fig.~\ref{fig:sweden_map} shows two practical train services operating on these corridors, denoted as Train~245 and Train~965, whose routes overlap in the Stockholm central region called SE3 and interact through shared substations. The main station and connection nodes include Uppsala~C, Arlanda~C, H\"aggvik, Stockholm~C, \"Alvsj\"o, Flemingsberg, S\"odert\"alje Syd, J\"arna, Nykvarn, Eskilstuna~C, and T\"alle, together with the southbound corridor towards Nyk\"oping~C. The corresponding train timetables and service information are obtained from the Trafikverket platform \cite{b17}.

\begin{table}[t]
\caption{Train profile for the Swedish case study.}
\label{tab:train_profile}
\centering
\begin{tabular}{p{3.2cm} p{4.7cm}}
\hline
\textbf{Item} & \textbf{Specification} \\
\hline
Service & M\"alart\aa g regional service \\
Operator & M\"alardalstrafik (SJ/MTR) \\
Train type & Regional EMU (CAF ER1 Regina) \\
Service role & Regional and commuter \\
Typical max speed & 200~km/h (typically 160~km/h in service) \\
Typical weight & $\sim$200~t (4-car set) \\
Train length & $\sim$100~m \\
Main routes & Stockholm, Uppsala, V\"aster\aa s, Eskilstuna \\
\hline
\end{tabular}
\end{table}

The trains are modeled as regional electric multiple units (EMUs) operating on the selected Swedish corridors. The operator and rolling stock type are summarized in Table~\ref{tab:train_profile}. For the multi-train movement model, the key parameters used to generate speed and trajectory profiles are given in Table~\ref{tab:train_params}. The effective mass is conservatively set to account for passenger load and rotary allowance. The regenerative braking efficiency is used to represent the recoverable percentage of braking energy.

\begin{table}[t]
\caption{Key parameters used in the train movement and power demand model.}
\label{tab:train_params}
\centering
\begin{tabular}{l c c}
\hline
\textbf{Parameter} & \textbf{Value} & \textbf{Unit} \\
\hline
Effective train mass $M'$ & 200{,}000 & kg \\
Gravitational acceleration $g$ & 9.81 & m/s$^{2}$ \\
Maximum speed $v_{\max}$ & 40 & m/s \\
Minimum speed $v_{\min}$ & 20 & m/s \\
Maximum acceleration $a_{\max}$ & 0.2 & m/s$^{2}$ \\
Maximum deceleration $a_{\min}$ & $-0.1$ & m/s$^{2}$ \\
Davis coefficient $A_h$ & 500 & N \\
Davis coefficient $B_h$ & 30 & N$\cdot$s/m \\
Davis coefficient $C_h$ & 5 & N$\cdot$s$^{2}$/m$^{2}$ \\
Regenerative efficiency $\eta_{\mathrm{reg}}$ & 0.85 &  \\
\hline
\end{tabular}
\end{table}

\begin{figure}[htbp]
\centerline{\includegraphics[width=0.45\textwidth]{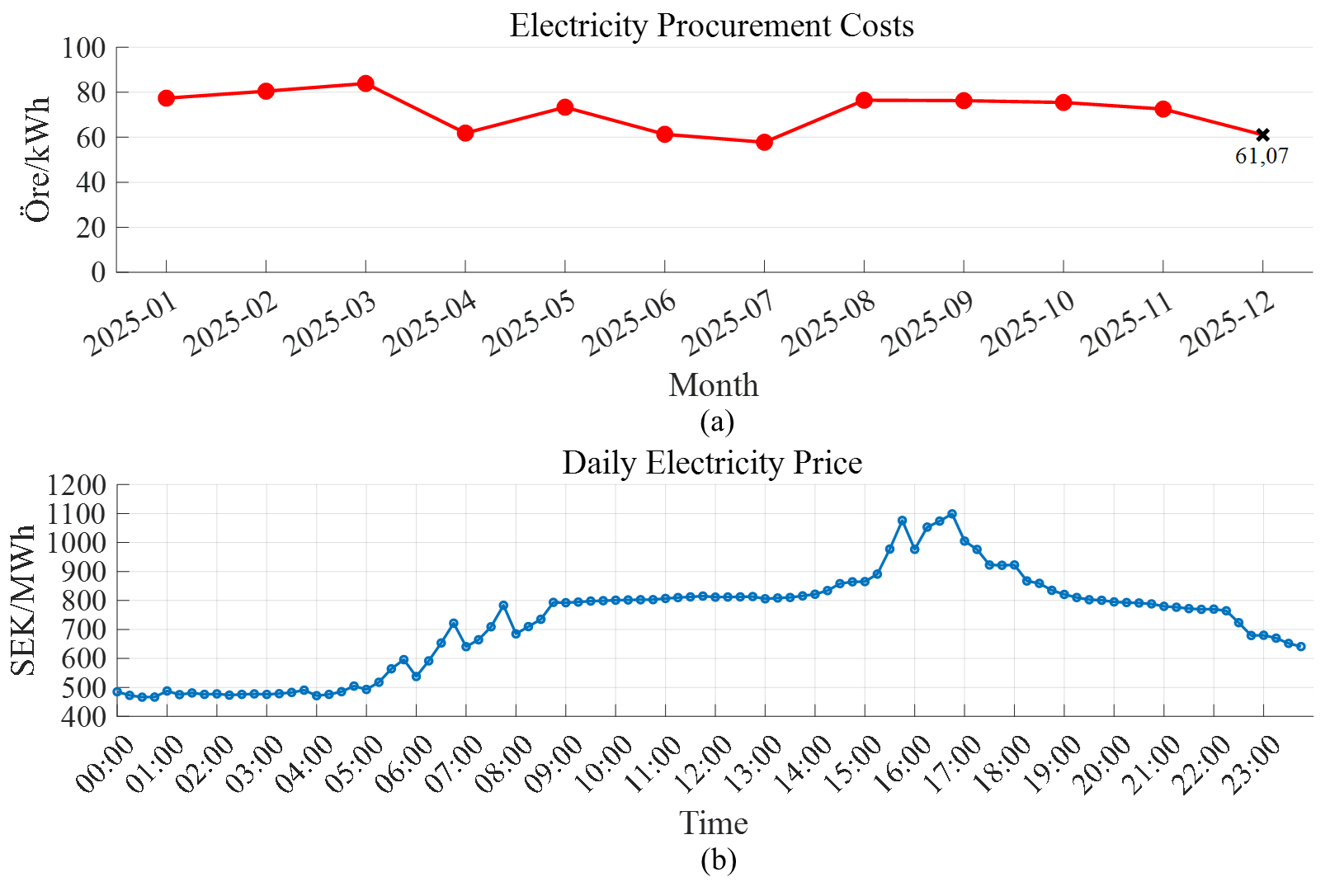}}
\caption{The electricity price: (a) monthly average procurement cost from Trafikverket in Sweden; (b) 24-hour electricity price.}
\label{fig:price_monthly}
\end{figure}

Electricity price data are collected in two layers. 
Firstly, the monthly electricity procurement price for railway traction is taken from Trafikverket's electricity price reports \cite{b18,b19}, as shown in Fig.~\ref{fig:price_monthly}(a). The reported values are in $\ddot{\mathrm{O}}$re/kWh, where $1~\ddot{\mathrm{O}}\mathrm{re}/\mathrm{kWh}=10~\mathrm{SEK}/\mathrm{MWh}$. The December value $61.07~\ddot{\mathrm{O}}\mathrm{re}/\mathrm{kWh}$ is adopted as the base price for scaling and normalization. Secondly, intra-day price variability is represented using an hourly price profile for the Swedish SE3 bidding zone, shown in Fig.~\ref{fig:price_monthly}(b). The normalized hourly fluctuations are used to construct the time-varying price for both the day-ahead operation and the intra-day rolling optimization.

The ESS economic parameters are derived from the International Renewable Energy Agency (IRENA) reference data \cite{b20}. The unit investment cost is selected from the 10~MW class cost entry, where the energy cost is about $1500$~SEK/kWh, the power cost is about $1700$~SEK/kW. The efficiency, lifetime, usable SOC range, and end-of-life capacity retention are summarized in Table~\ref{tab:ess_params} and used to parameterize the investment model and the SOC dynamics in the optimization.

\begin{table}[t]
\caption{ESS economic parameters used in the case study.}
\label{tab:ess_params}
\centering
\begin{tabular}{l c c}
\hline
\textbf{Parameter} & \textbf{Value} & \textbf{Unit} \\
\hline
Technology & Li-ion &  \\
Total installation cost & 1920 & SEK/kWh \\
Energy cost $C_e$ & 1500 & SEK/kWh \\
Power cost $C_p$ & 1700 & SEK/kW \\
Round-trip efficiency & 85\%--90\% &  \\
Usable SOC range & 80\%--100\% &  \\
Lifetime & 10 & years \\
Capacity at end of life & 70\%--80\% & \\
O\&M cost ratio & 2\% &  \\
\hline
\end{tabular}
\end{table}

Four cases are constructed to evaluate the incremental benefits of ESS integration, trajectory optimization, and intra-day adjustment. Base case shows the RPSS operation with the initial train trajectories and without any ESS installation. In case 1, ESSs are integrated while the train trajectories remain unchanged, so that the impact of ESS siting, sizing, and operation can be isolated under the same baseline. In case 2, both ESS integration and train trajectory optimization are enabled, where the trajectory candidates are iteratively selected and the resulting train power profiles are fed into the day-ahead operation to co-optimize the railway operation and the ESS deployment. In case 3, an uncertainty-aware intra-day rolling optimization is further applied to correct short term deviations caused by PV and load forecast errors.

\subsection{Base case: Initial train trajectory, without ESS}

\begin{figure}[htbp]
\centerline{\includegraphics[width=0.4\textwidth]{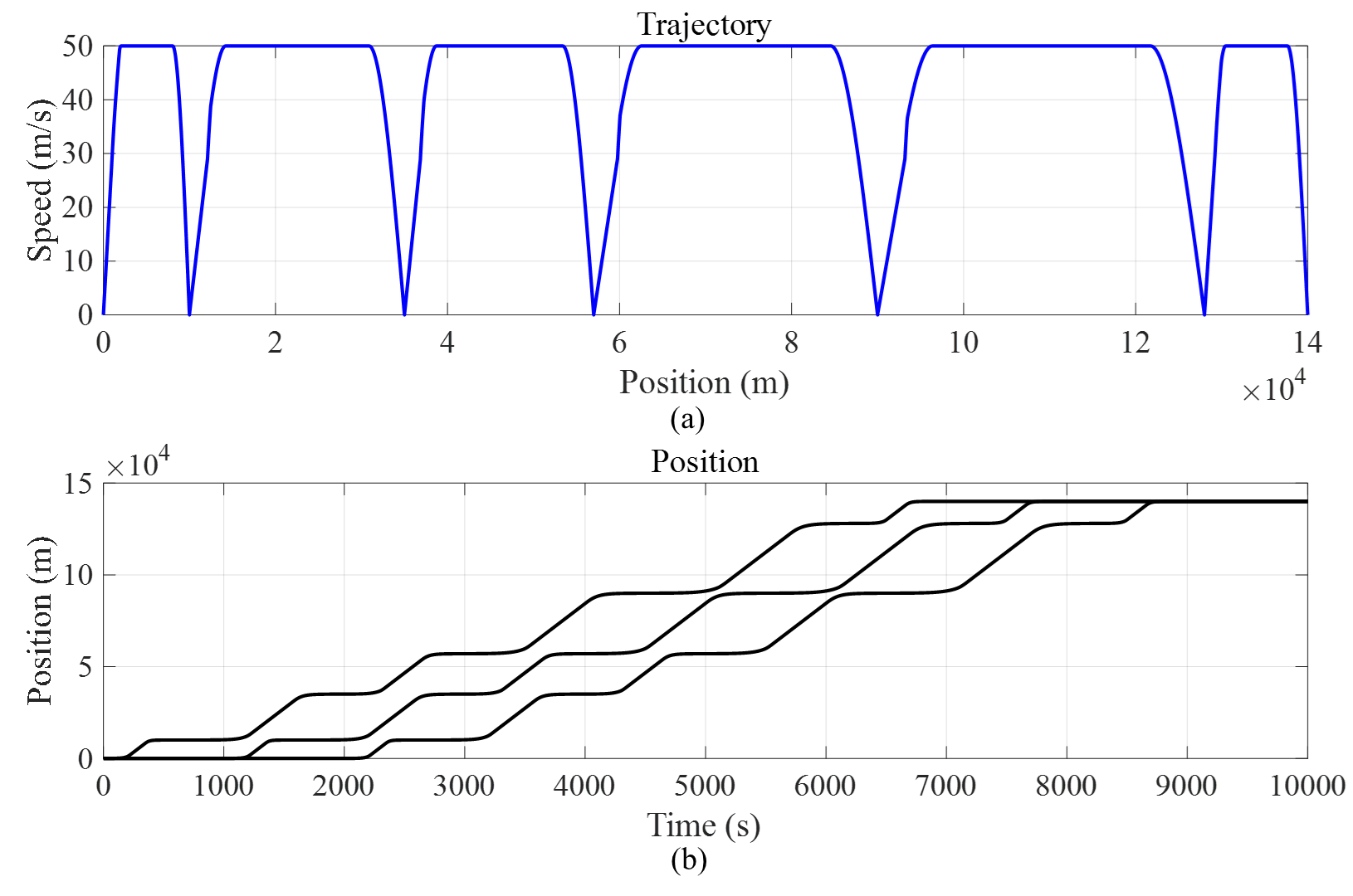}}
\caption{The movement profile of Train 245 in base case: (a) Trajectory; (b) Position.}
\label{fig:trajectory_initial_245}
\end{figure}

\begin{figure}[htbp]
\centerline{\includegraphics[width=0.4\textwidth]{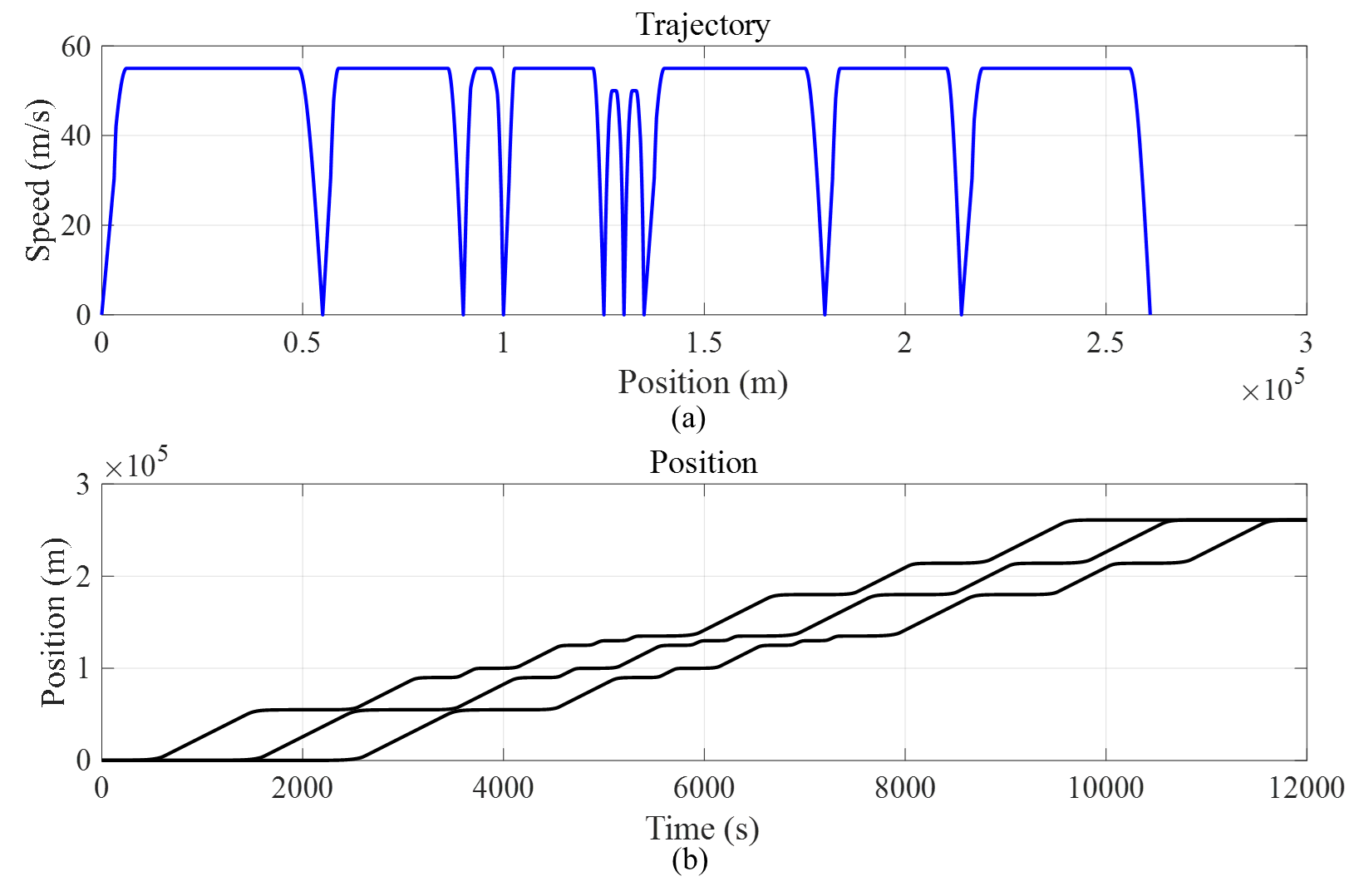}}
\caption{The movement profile of Train 965 in base case: (a) Trajectory; (b) Position.}
\label{fig:trajectory_initial_965}
\end{figure}

Under the initial trajectories, the movement profiles in Fig.~\ref{fig:trajectory_initial_245} and Fig.~\ref{fig:trajectory_initial_965} exhibit the typical four phase operating pattern, acceleration, cruising, deceleration, and dwelling. The computed traction force is positive during acceleration and cruising, while it becomes negative during braking, indicating regenerative operation. Accordingly, the baseline active power demand presents pronounced peaks during acceleration segments and partial power recovery during braking segments.

\begin{table}[t]
\caption{Energy recovery in the base case.}
\label{tab:base_regen_benefit}
\centering
\begin{tabular}{l c c c}
\hline
\textbf{Metric} & \textbf{Symbol} & \textbf{Train 245} & \textbf{Train 965} \\
\hline
Peak net power & $P_{\mathrm{peak}}$ (MW) & 7.8 & 9.9 \\
Net grid energy & $E_{\mathrm{grid}}$ (MWh) & 0.6 & 1.3 \\
Traction energy & $E_{\mathrm{traction}}$ (MWh) & 0.9 & 1.8 \\
Braking energy & $E_{\mathrm{regen}}$ (MWh) & 0.3 & 0.6 \\
Potential regenerative ratio & $R_{\mathrm{regen}}$ (\%) & 33.4 & 30.0 \\
\hline
\end{tabular}
\end{table}

\begin{figure}[htbp]
\centerline{\includegraphics[width=0.4\textwidth]{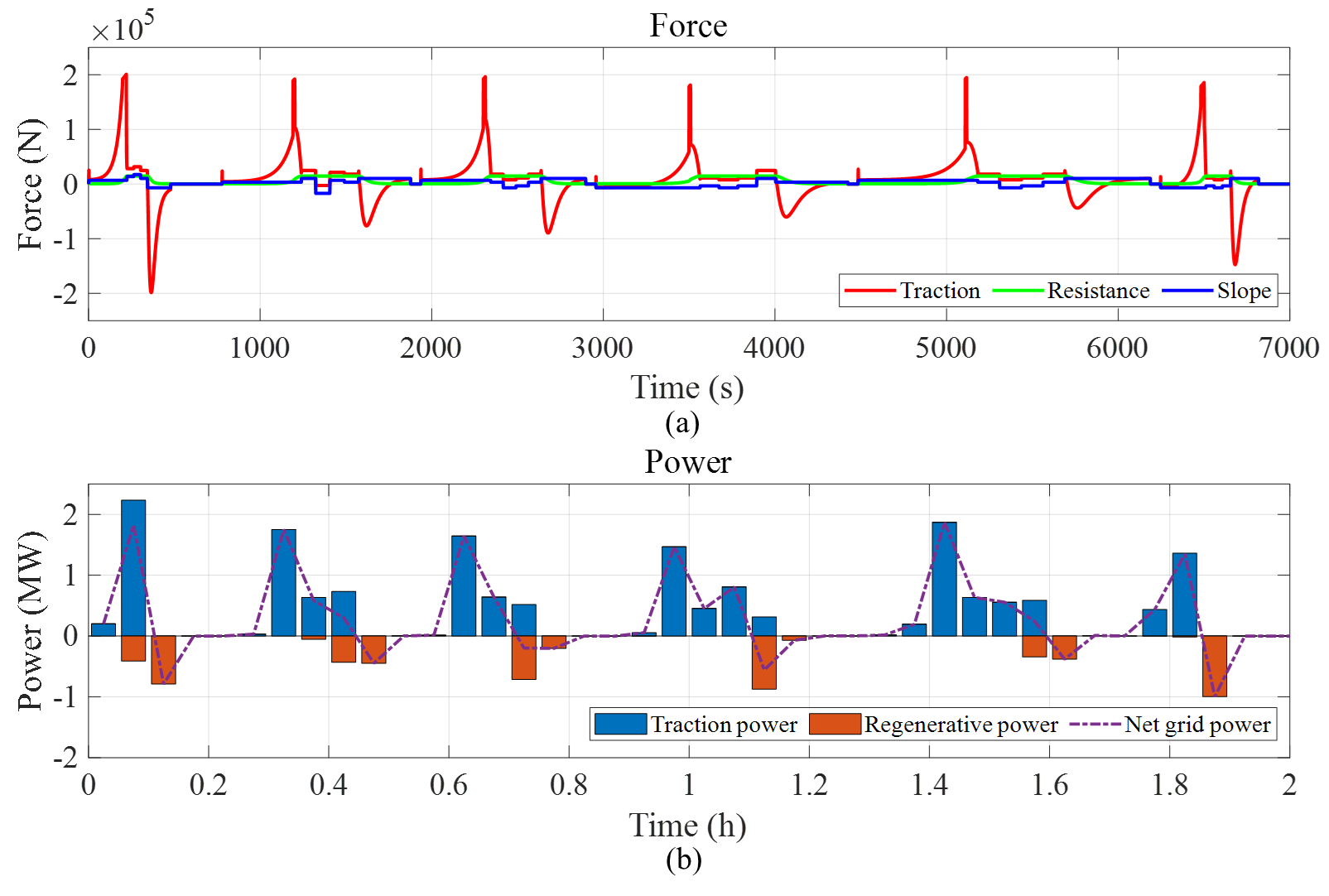}}
\caption{The different types of forces and power demand of Train 245 in base case: (a) Force; (b) Power.}
\label{fig:powerdemand_initial_245}
\end{figure}

\begin{figure}[htbp]
\centerline{\includegraphics[width=0.4\textwidth]{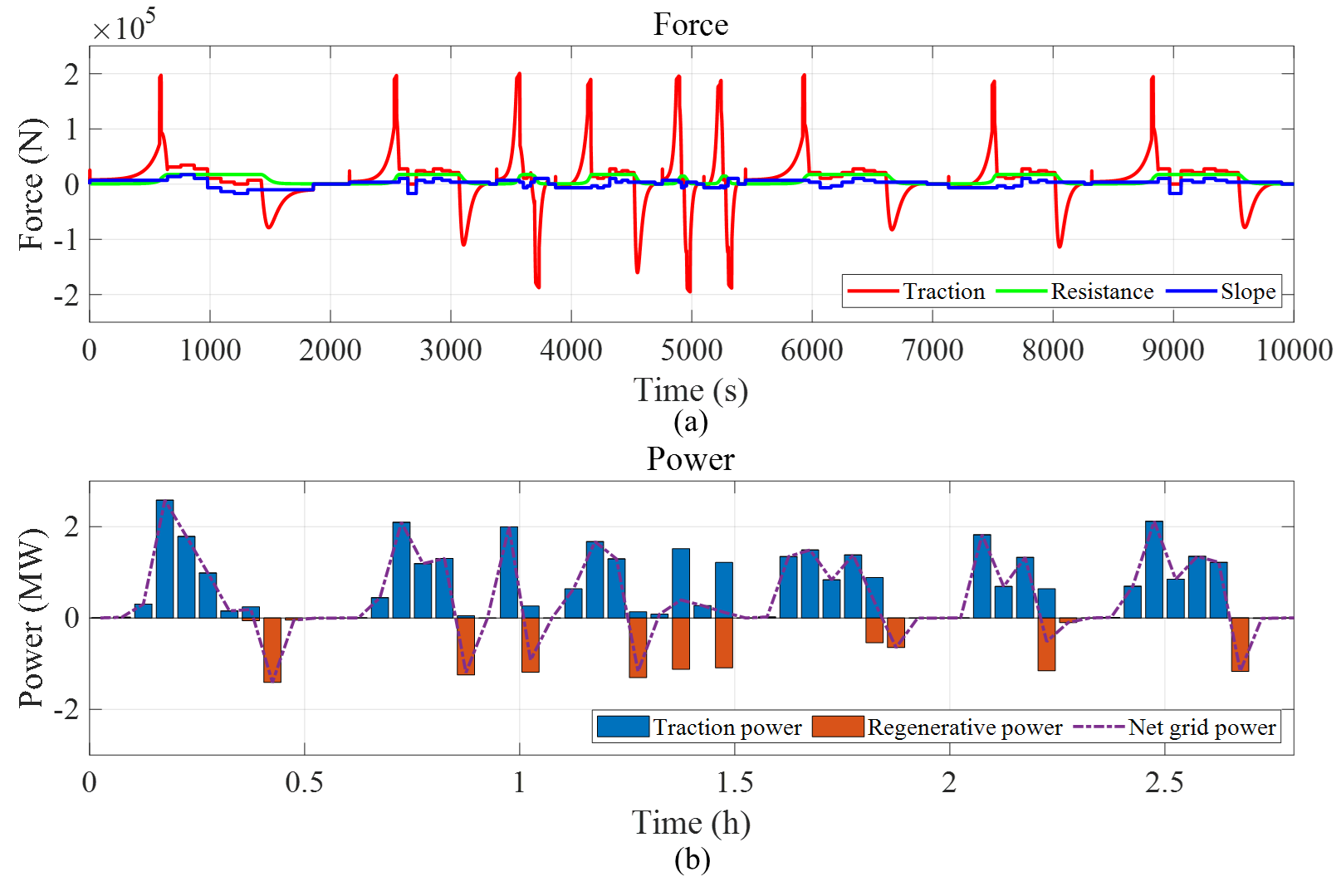}}
\caption{The different types of forces and power demand of Train 965 in base case: (a) Force; (b) Power.}
\label{fig:powerdemand_initial_965}
\end{figure}

The force and power results in Fig.~\ref{fig:powerdemand_initial_245} and Fig.~\ref{fig:powerdemand_initial_965} reveal a consistent physical mechanism behind the baseline demand. 
Table~\ref{tab:base_regen_benefit} summarizes the potentially regenerative braking energy in the base case. If the braking energy could be fully captured and reused, the regenerative ratio would reach about 33\% for Train~245 and 30\% for Train~965, corresponding to 0.3~MWh and 0.6~MWh, respectively.
However, acceleration driven power spikes dominate the instantaneous grid exchange, whereas regenerative intervals offset the preceding traction demand only partially because braking and traction demands rarely coincide in time and location. With ESS integration, a larger share of the available braking energy can be stored and shifted to subsequent traction intervals, leading to higher energy utilization efficiency. Therefore, from the grid interaction perspective, the baseline $P_{\mathrm{grid}}(t)$ profile features pronounced peaks and relatively high short term variability. These baseline characteristics are used as the benchmark in Case~1 to evaluate pure ESS integration under fixed trajectories, and in Case~2 to further evaluate the benefit of trajectory smoothing on peak reduction.

\subsection{Case 1: Initial Trajectories With ESS Integration}

In Case~1, the initial trajectories of Train~245 and Train~965 are kept unchanged. Three ESS units are installed at
 \"Alvsj\"o, Eskilstuna~C, and Nyk\"oping~C. The day-ahead operation determines the ESS sizing and dispatch with the objective of peak reduction. Based on the selected configuration, the ESS energy capacities are optimized to $E^{\max}=[0.4,\,0.35,\,0.25]$~MWh and the charging and discharging power limits are $P^{\max}=[4.6,\,3.8,\,2.5]$~MW.

\begin{figure}[htbp]
\centerline{\includegraphics[width=0.45\textwidth]{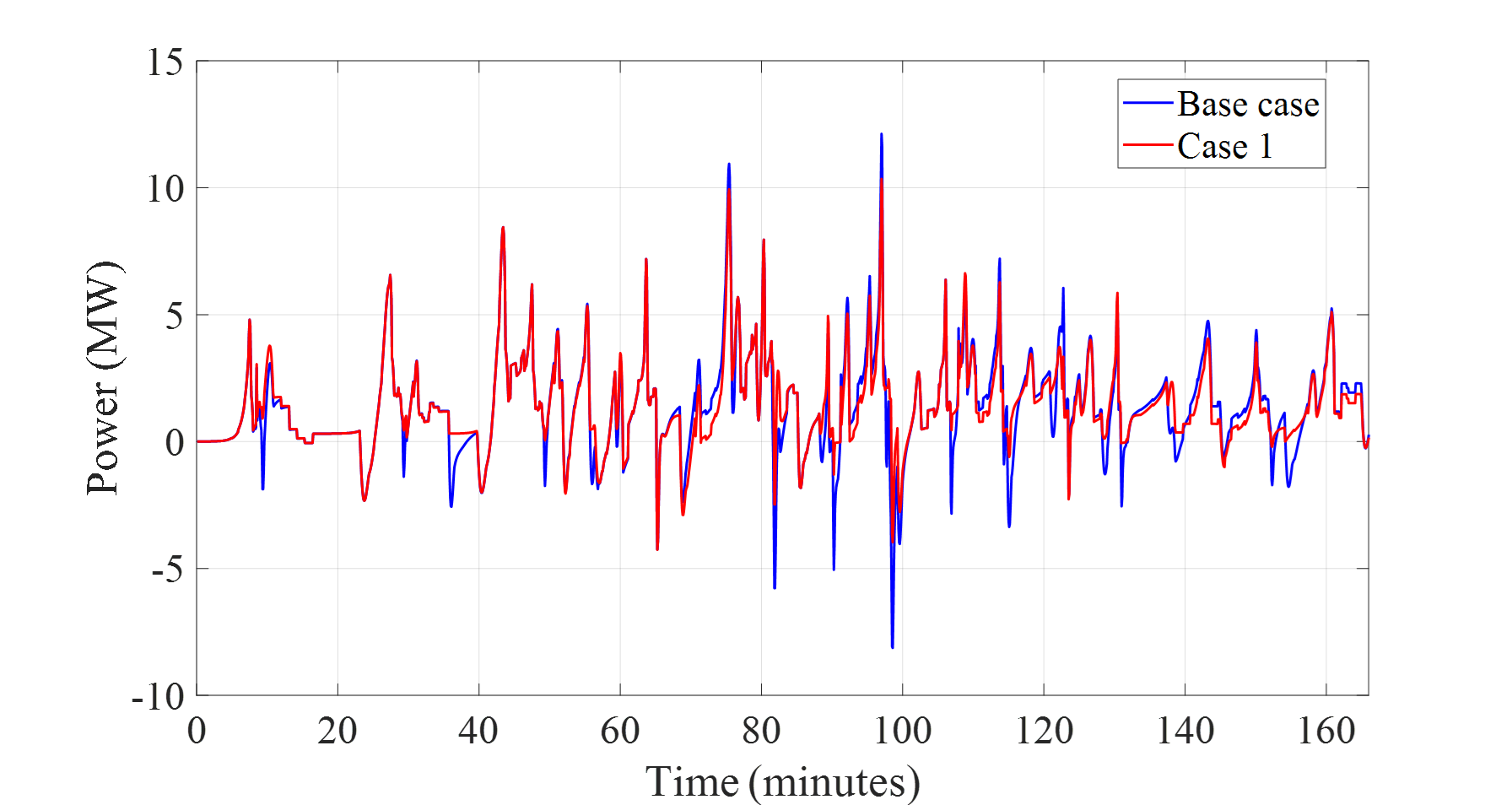}}
\caption{Grid power demand comparison between base case and case 1.}
\label{fig:power_case1}
\end{figure}

\begin{figure}[htbp]
\centerline{\includegraphics[width=0.45\textwidth]{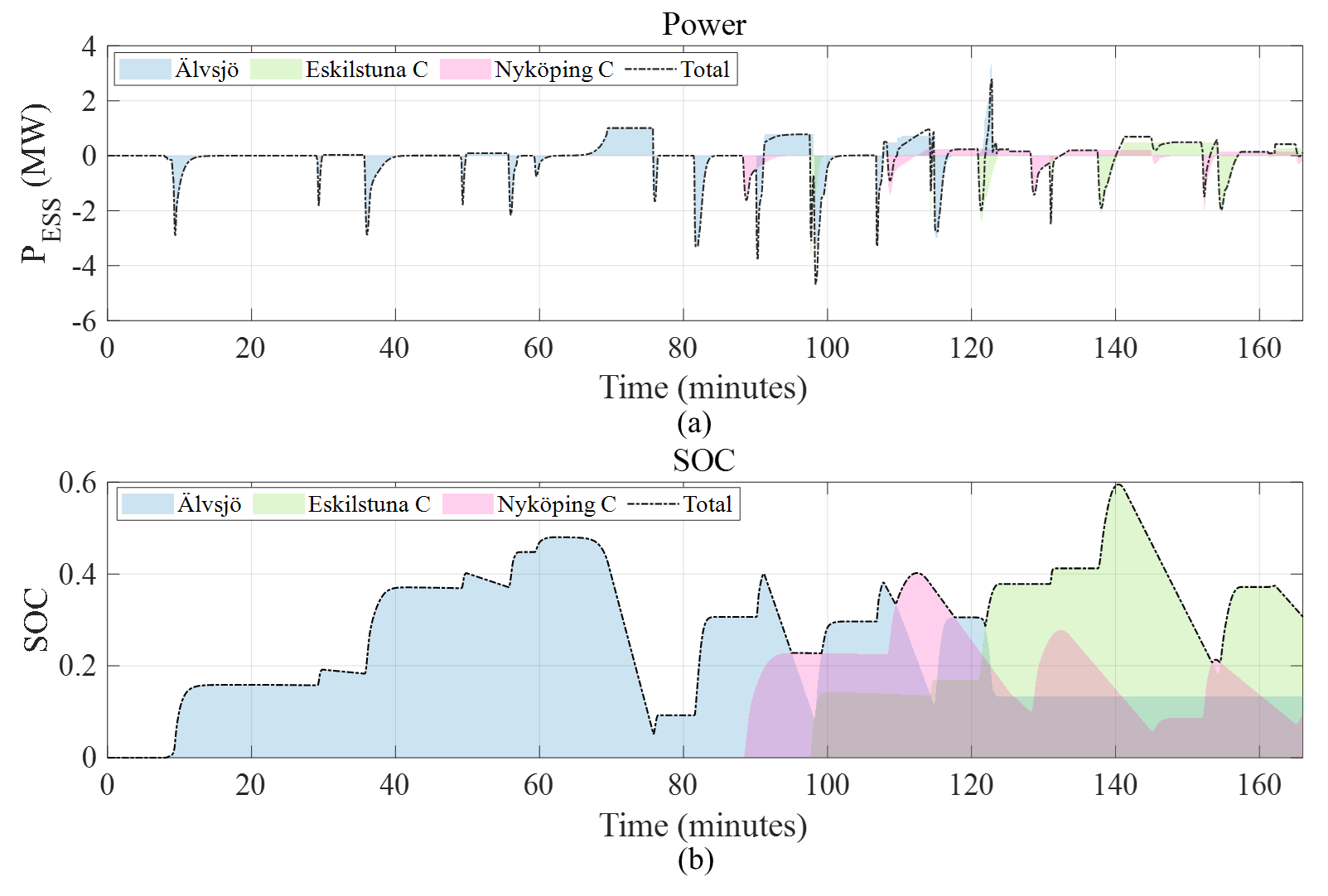}}
\caption{The ESS power output and SOC in case 1: (a)ESS power outout; (b)ESS SOC}
\label{fig:ESS_case1}
\end{figure}

Fig.~\ref{fig:power_case1} compares the baseline grid exchange profile and the Case~1 result after ESS operation. The ESS operating behavior is illustrated in Fig.~\ref{fig:ESS_case1}. The ESS power output shows frequent short pulses, indicating that the units respond primarily to transient events. This pattern leads to a lower peak level and a smoother grid power demand profile. Quantitatively, the peak grid demand is reduced from $12.1$~MW to $10.4$~MW, corresponding to a $14.7\%$ reduction. The grid energy decreases from $4.9$~MWh to $4.6$~MWh, yielding an energy saving of $5.9\%$. 
From the economic perspective, the reduced peak and energy import translate into a lower grid purchase cost. The grid side cost decreases from $14740.7$~SEK in the base case to $12968.8$~SEK in Case~1, corresponding to a $12.0\%$ reduction. When the ESS cost is allocated, the total cost equals to $14394.6$~SEK, resulting in a net cost improvement of $2.4\%$ for the whole period. This motivates Case~2, where train trajectories are further optimized to align traction demand.

\subsection{Case 2: Optimized Trajectories With ESS Integration}

\begin{figure}[htbp]
\centerline{\includegraphics[width=0.4\textwidth]{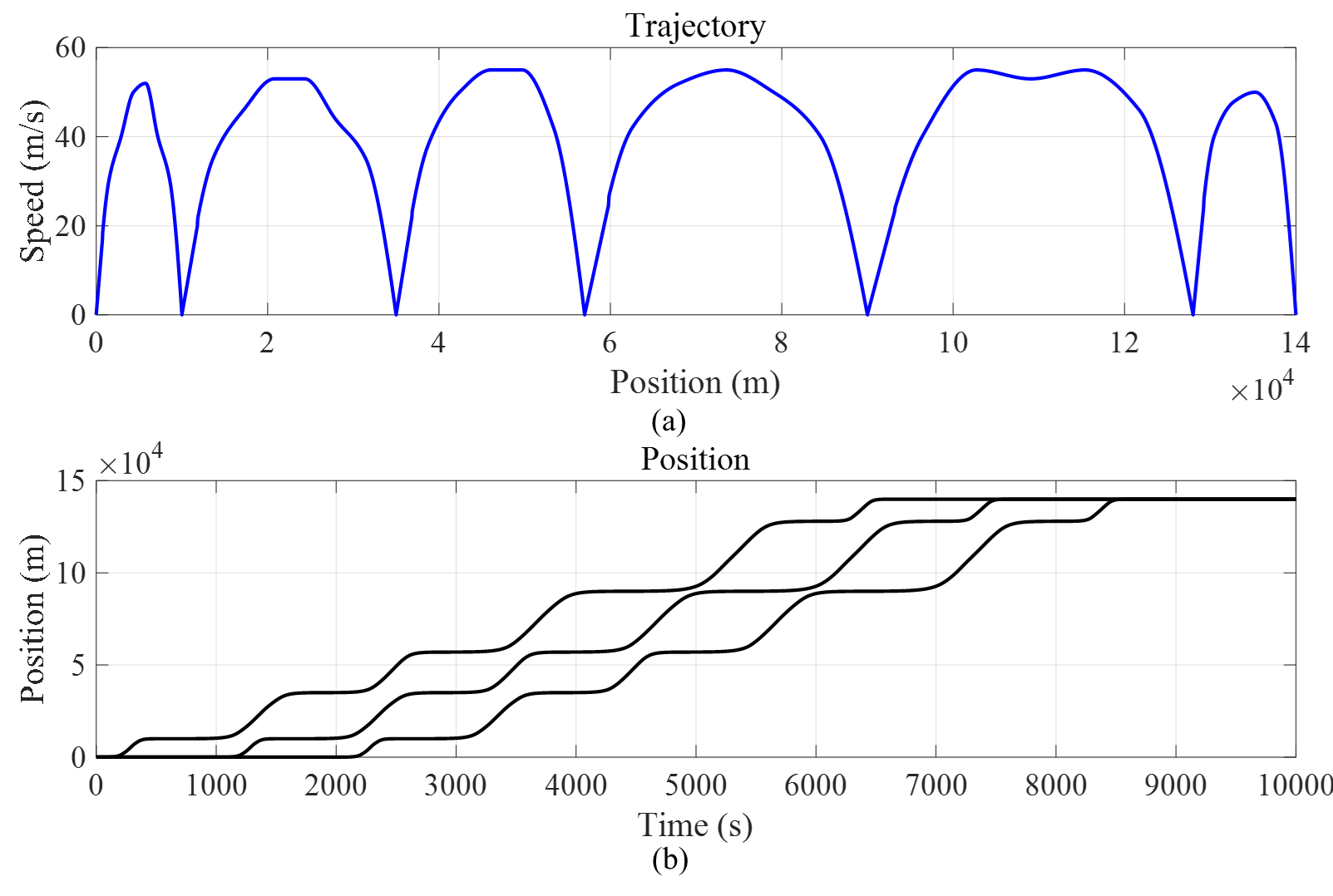}}
\caption{The movement profile of Train 245 in case 2: (a) Trajectory; (b) Position.}
\label{fig:trajectory_case2_245}
\end{figure}

\begin{figure}[htbp]
\centerline{\includegraphics[width=0.4\textwidth]{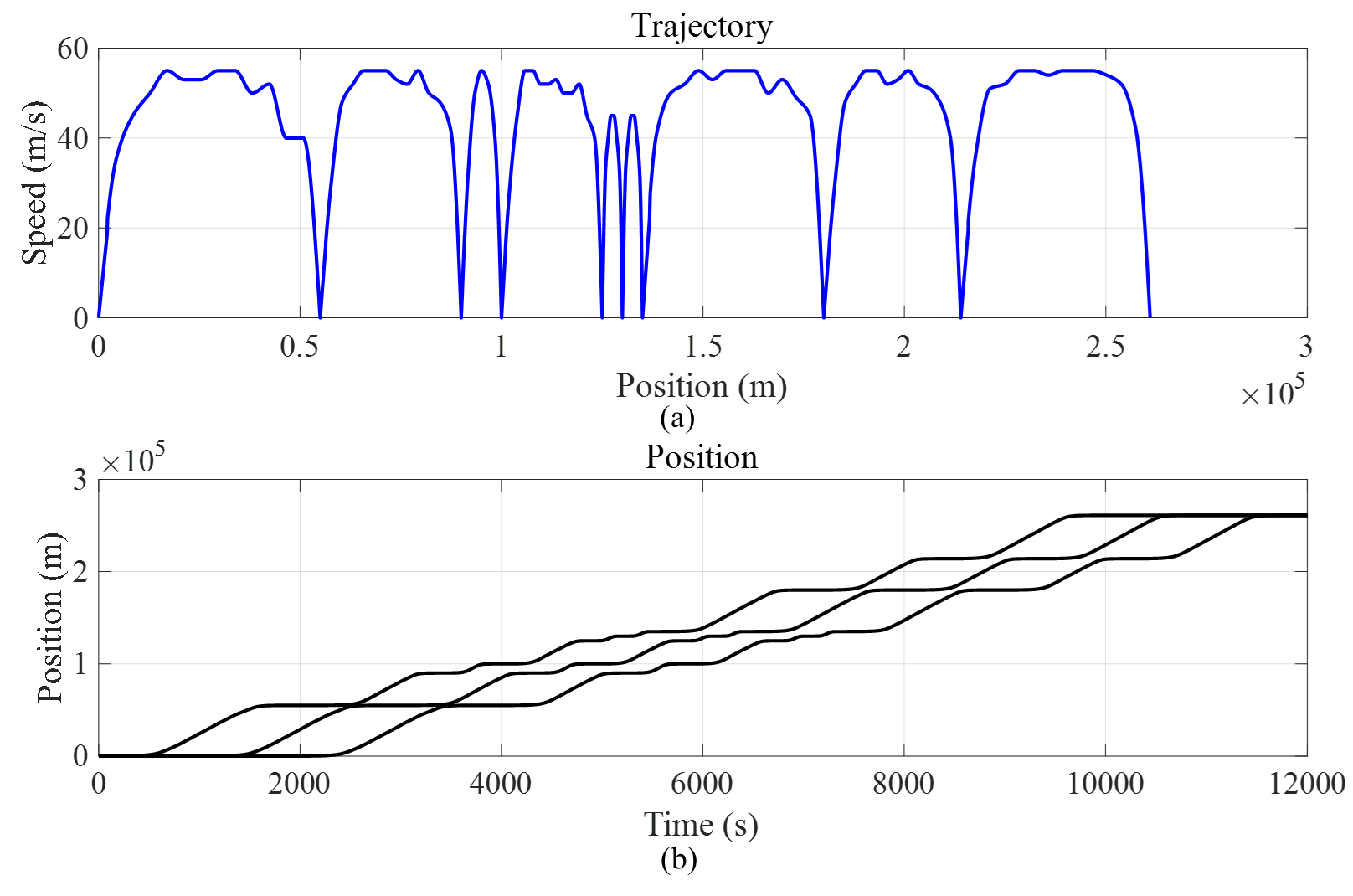}}
\caption{The movement profile of Train 965 in case 2: (a) Trajectory; (b) Position.}
\label{fig:trajectory_case2_965}
\end{figure}

Case~2 enables both ESS integration and trajectory optimization. Candidate trajectories are generated and evaluated iteratively for the day-ahead operation. Compared with the base case, the optimized trajectories in Fig.~\ref{fig:trajectory_case2_245} and Fig.~\ref{fig:trajectory_case2_965} show smoother speed transitions, where acceleration and braking events are redistributed in time to avoid excessive clustering. The position curves indicate that the optimized operation preserves the service pattern while adjusting the timing of inter station runs, which increases flexibility for coordinating multi-train demand on shared supply buses.

\begin{figure}[htbp]
\centerline{\includegraphics[width=0.4\textwidth]{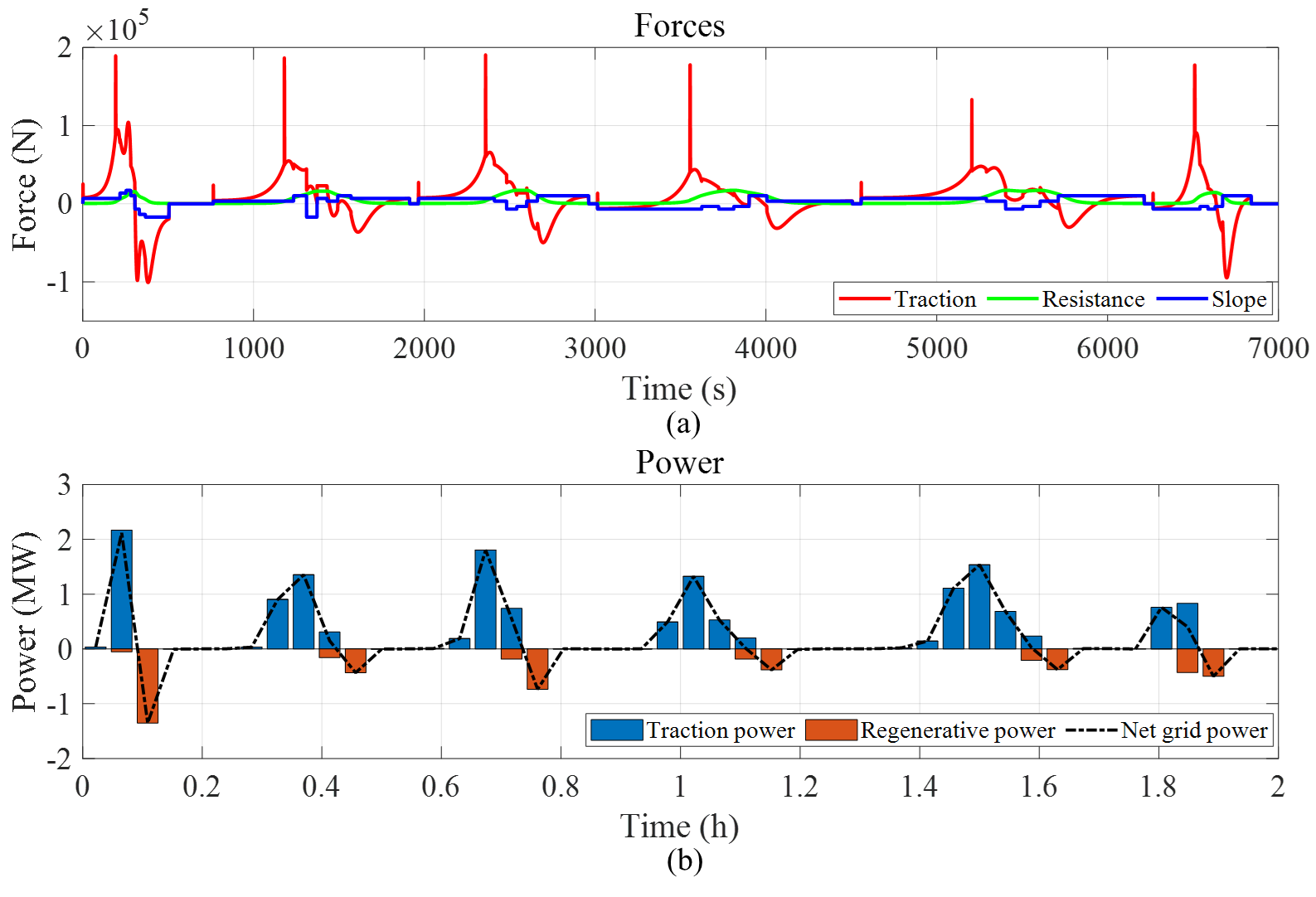}}
\caption{The different types of forces and power demand of Train 245 in case 2: (a) Force; (b) Power.}
\label{fig:powerdemand_case2_245}
\end{figure}

\begin{figure}[htbp]
\centerline{\includegraphics[width=0.4\textwidth]{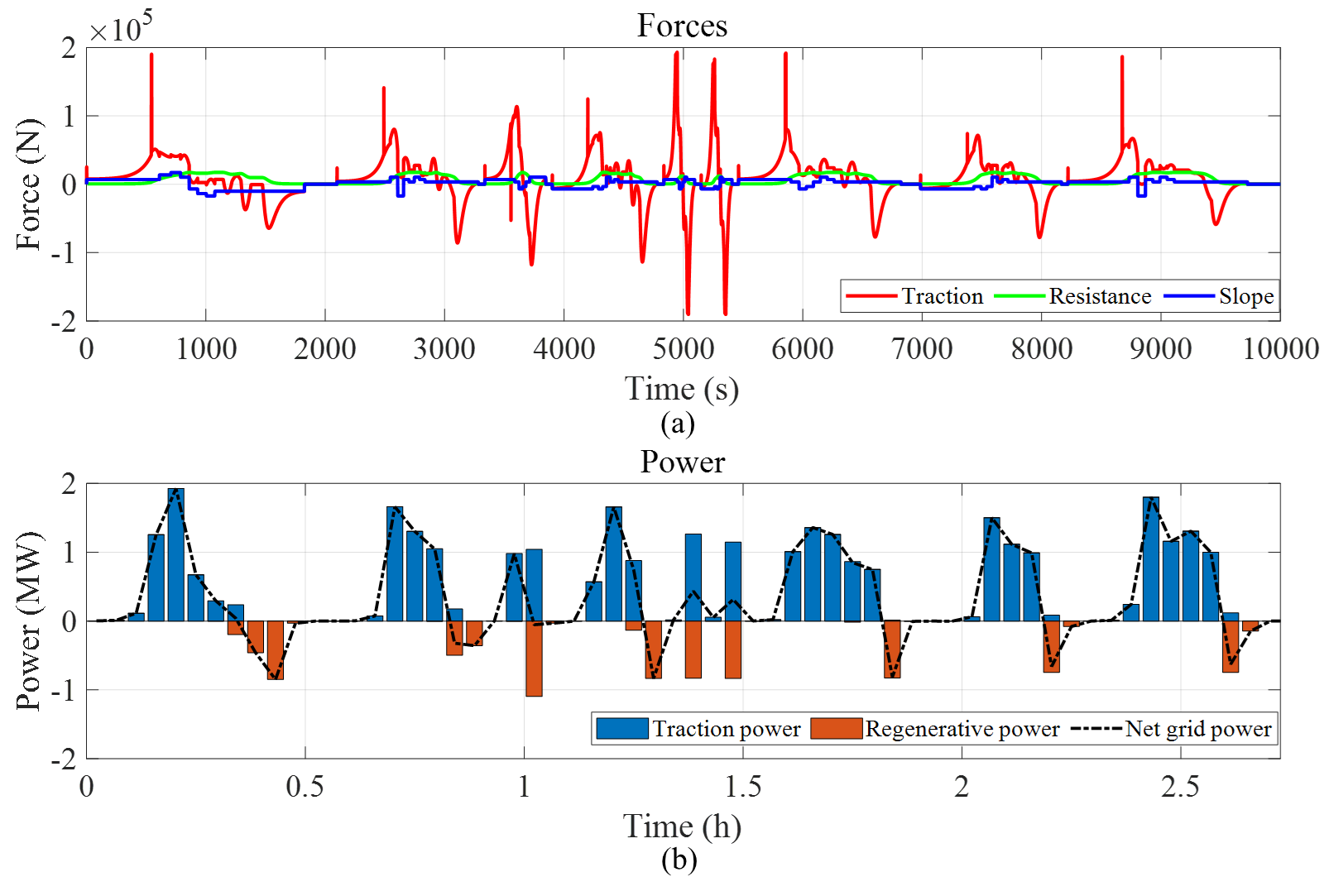}}
\caption{The different types of forces and power demand of Train 965 in case 2: (a) Force; (b) Power.}
\label{fig:powerdemand_case2_965}
\end{figure}

\begin{figure}[htbp]
\centerline{\includegraphics[width=0.45\textwidth]{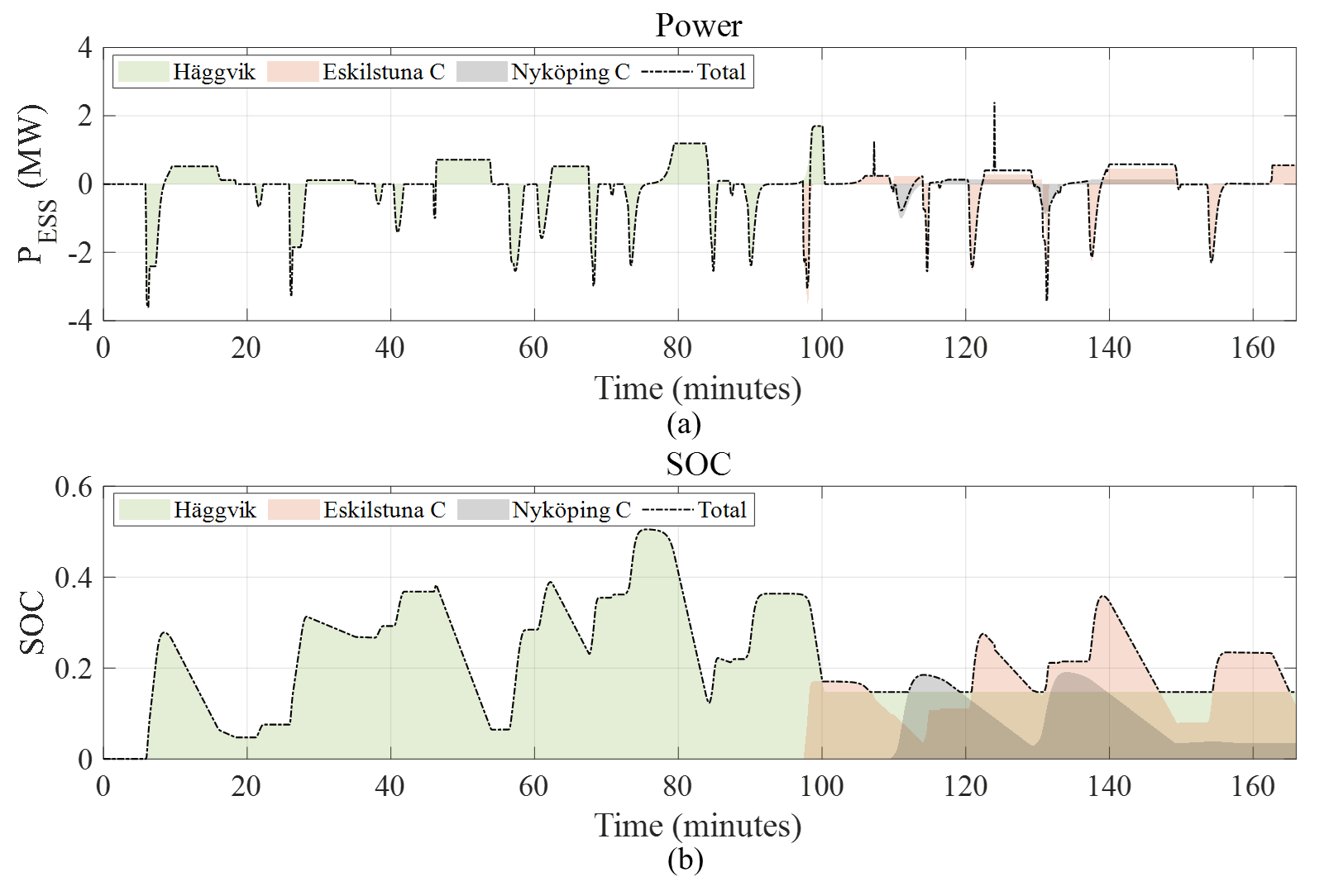}}
\caption{The ESS power output and SOC in case 2: (a)ESS power outout; (b)ESS SOC}
\label{fig:ESS_case2}
\end{figure}

The force and power profiles are shown in Fig.~\ref{fig:powerdemand_case2_245} and Fig.~\ref{fig:powerdemand_case2_965}. Compared with case~1, the traction power blocks become less concentrated and the grid power exhibits fewer sharp spikes. 
In this case, three ESS units are deployed at H\"aggvik, Eskilstuna~C, and Nyk\"oping~C. The selected sizing is $E^{\max}=[0.34,\,0.28,\,0.29]$~MWh with charging power limits $P^{\max}=[3.6,\,3.5,\,3.2]$~MW. Under the optimized trajectories, the peak grid demand is reduced to $8.9$~MW, and the net grid energy decreases to $4.2$~MWh. The ESS power output and SOC in Fig.~\ref{fig:ESS_case2} indicate more effective charging during braking dominant periods and more targeted discharging around acceleration dominant periods.

The improvement in Case~2 comes from two coupled mechanisms. Trajectory smoothing reduces the formation of concurrent acceleration peaks, and the redistributed braking events create charging opportunities that are better aligned with traction demand. As a result, the ESS dispatch is no longer dominated by highly concentrated spikes and can provide peak mitigation with less aggressive short term response.

\subsection{Case 3: Optimized Trajectories and ESS With Uncertainty Consideration}

Case~3 incorporates uncertainty and applies a two stage rolling optimization for intra-day correction. The optimized trajectories and ESS deployment obtained in Stage~I are retained as the planned operation, while Stage~II updates the ESS dispatch in a rolling horizon manner using updated forecasts of traction demand and PV output. PV units are connected at \AA lvsj\"o, \"Orebro~C, and Norrk\"oping~C, and the rolling adjustment aims to reduce peak excursions and improve the tracking of the planned grid exchange under forecast errors.

\begin{figure}[htbp]
\centerline{\includegraphics[width=0.45\textwidth]{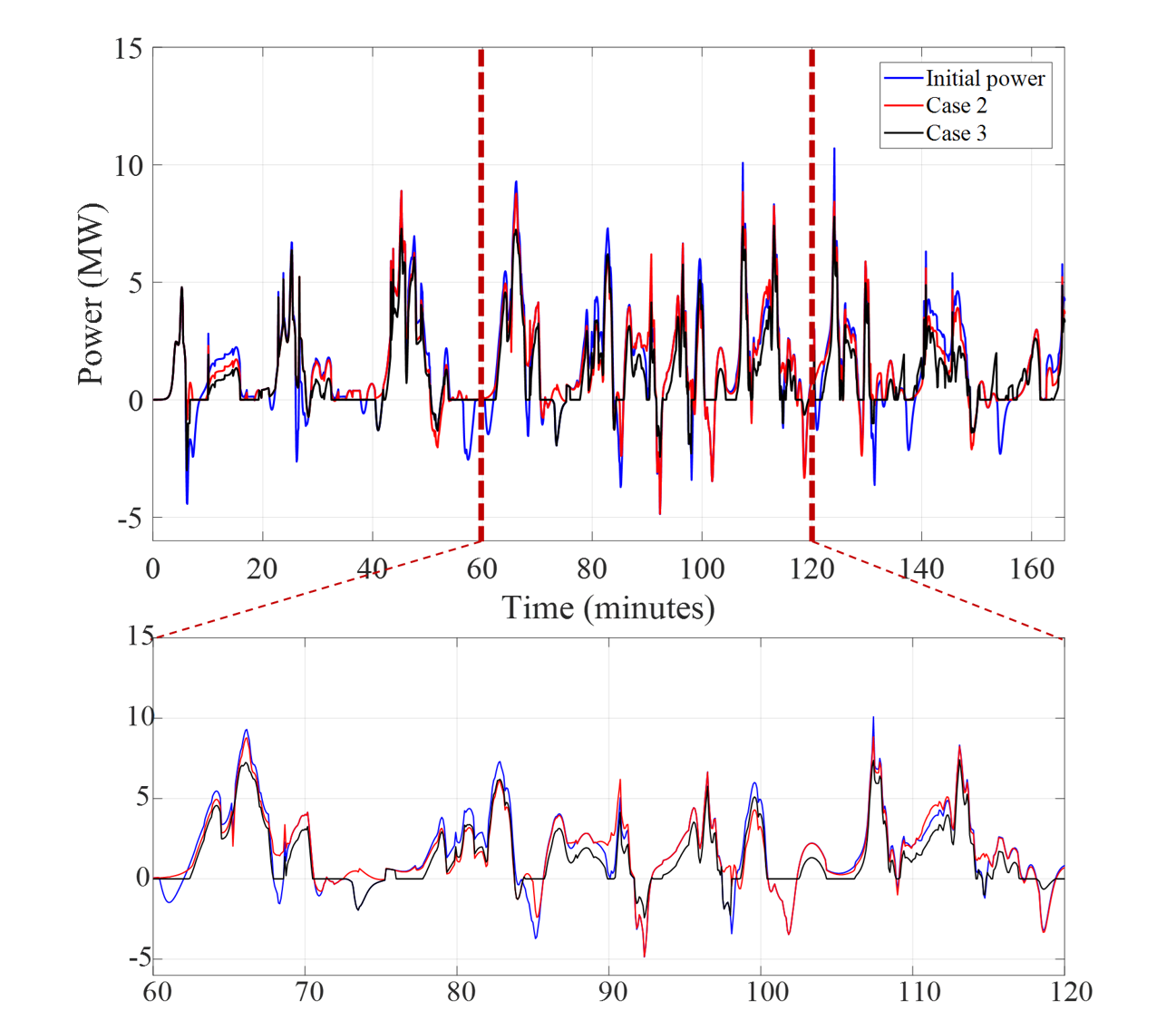}}
\caption{Overall grid power demand comparison.}
\label{fig:power_case3}
\end{figure}

Figure~\ref{fig:power_case3} compares three grid power profiles. The curve labeled Initial power represents the planned grid exchange associated with the optimized trajectories from Case~2 before applying the rolling correction. It therefore differs from the base case defined by the initial trajectories. The Case~2 curve corresponds to the day-ahead result with optimized trajectories and ESS integration. The Case~3 curve shows the realized profile after the rolling adjustment with PV uncertainty. By updating the dispatch using refreshed forecasts, the ESS charging and discharging actions are shifted in time to compensate short term deviations and suppress transient peaks that are not fully anticipated in the day-ahead plan. In parallel, PV injections reduce the net grid import at the connected buses during periods with available generation, which contributes to lower grid energy and purchase cost.

As a result, Case~3 achieves a peak grid demand of $7.8$~MW and a net grid energy of $3.2$~MWh. The corresponding grid purchase cost is $9533.1$~SEK. When the allocated ESS cost is included, the total cost becomes $10494.6$~SEK. Table~\ref{tab:overall_comparison} provides a consolidated comparison across the base case and Cases~1 to~3.

\begin{table*}[t]
\caption{Overall comparison across the base case and cases 1 to 3.}
\label{tab:overall_comparison}
\centering
\begin{tabular}{l c c c c c c c}
\hline
\textbf{Case} & \textbf{$P_{\mathrm{peak}}$} & \textbf{Peak saving} & \textbf{$E_{\mathrm{grid}}$} & \textbf{$C_{\mathrm{grid}}$} & \textbf{$C_{\mathrm{ESS}}$} & \textbf{$C_{\mathrm{total}}$} & \textbf{Total saving} \\
 & \textbf{MW} & \textbf{\%} & \textbf{MWh} & \textbf{SEK} & \textbf{SEK} & \textbf{SEK} & \textbf{\%} \\
\hline
Base case & 12.1 & 0.0 & 4.9 & 14740.7 & 0.0 & 14740.7 & 0.0 \\
Case 1 & 10.4 & 14.7 & 4.7 & 12968.8 & 1425.8 & 14394.6 & 2.4 \\
Case 2 & 8.9 & 26.7 & 4.2 & 11349.5 & 1137.4 & 12486.9 & 15.3 \\
Case 3 & 7.8 & 35.7 & 3.2 & 9533.1 & 961.4 & 10494.6 & 28.8 \\
\hline
\end{tabular}
\end{table*}

\section{Conclusion}

This paper proposes a two-stage coordinated optimization model for RPSS energy management, where day-ahead planning co-optimizes train trajectories with ESS siting and sizing, and an intra-day rolling stage based on AWC-MPC updates ESS dispatch using refreshed forecasts of traction demand and PV output. Case studies show that the proposed method realizes trajectory smoothing and coordinated ESS charging and discharging, which reshape the aggregate traction demand. With rolling updates under forecast deviations, the dispatch remains effective in suppressing residual short term peaks and maintaining consistent grid exchange performance. The real Swedish railway system achieves 35.7\% peak reduction and 28.8\% total cost reduction, which demonstrates practical implementation under realistic operating conditions.

\end{document}